\documentclass[aps,prb,preprint,superscriptaddress,endfloats*]{revtex4-1}

\usepackage{amsmath,amssymb}
\usepackage{graphicx}
\usepackage{color}
\usepackage{bm}

\usepackage{txfonts}

\renewcommand{\vec}[1]{\boldsymbol{#1}}
\newcommand{\mat}[1]{\mathbf{#1}}

\renewcommand{\tilde}[1]{\widetilde{#1}}

\begin{document}

\preprint{Preprint: Tsukamoto and Ono}

\title{Improvement of accuracy of wave-function-matching method for transport calculation}


\author{Shigeru Tsukamoto}
\email{s.tsukamoto@fz-juelich.de}
\affiliation{Peter Gr\"{u}nberg Institut \& Institute for Advanced Simulation, Forschungszentrum J\"{u}lich and JARA, D-52425 J\"{u}lich, Germany}

\author{Tomoya Ono}
\email{ono@ccs.tsukuba.ac.jp}
\affiliation{Center for Computational Sciences, University of Tsukuba, Tsukuba, Ibaraki 305-8577, Japan}

\author{Stefan Bl{\"u}gel}
\affiliation{Peter Gr\"{u}nberg Institut \& Institute for Advanced Simulation, Forschungszentrum J\"{u}lich and JARA, D-52425 J\"{u}lich, Germany}

\date{\today}

\begin{abstract}
The wave-function-matching (WFM) technique for first-principles transport-property calculations was modified by S\o{}rensen {\it et al.} so as to exclude rapidly decreasing evanescent waves [S\o{}rensen {\it et al.}, Phys. Rev. B {\bf 77}, 155301 (2008)]. However, this method lacks translational invariance of the transmission probability with respect to insertion of matching planes and consistency between the sum of the transmission and reflection probabilities and the number of channels in the transition region. We reformulate the WFM method since the original methods are formulated to include all the generalized Bloch waves. It is found that the translational invariance is destroyed by the overlap of the layers between the electrode and transition regions and by the pseudoinverses used to exclude the rapidly decreasing evanescent waves. We then devise a method that removes the overlap and calculates the transmission probability without the pseudoinverses. As a result, we find that the translational invariance of the transmission probability with respect to insertion of the extra layers is properly retained and the sum of the transmission and reflection probabilities exactly agrees with the number of channels. In addition, we prove that the accuracy in the transmission probability of this WFM technique is comparable with that obtained by the nonequilibrium Green's function method. Furthermore, we carry out the electron transport calculations on two-dimensional graphene sheets embedded with B--N line defects sandwiched between a pair of semi-infinite graphene electrodes and find the dependence of the electron transmission on the transverse momentum perpendicular to the direction of transport.
\end{abstract}


\maketitle

\section{Introduction}
\label{sec:introduction}
The study of electron transport in nanoscale systems is becoming important as the miniaturization of electronic devices proceeds because they are expected to exhibit considerably different transport properties from those of classical conductors. Owing to the complexity of the problem, such studies are strongly dependent on the existence of reliable numerical treatments. A number of numerical methods for calculating the electron-transport properties of nanoscale systems have been proposed so far, and some of them are combined with first-principles calculations. The currently used methods in the first-principles calculations are roughly categorized into two approaches. One approach uses the nonequilibrium Green's function.\cite{PhysRevB_65_165401,PhysRevB_59_11936,PhysRevB_63_245407,PhysRevB_77_155301} The relation between the conductance and Green's function has been derived within the nonequilibrium Keldysh formalism,\cite{keldysh} and the charge density in the equilibrium regime of energy is easily computed with the energy of a nonreal number. The other approach is the wave-function-matching (WFM) method, which computes the scattering wave functions (SWFs) providing a direct real-space picture of the scattering process.\cite{PhysRevB_44_8017,PhysRevB_52_005335,PhysRevB_51_005278,PhysRevB_59_002267,PhysRevB_66_161402,PhysRevB_67_195315,icp,PhysRevB_70_195402,PhysRevB_86_195406,PhysRevB_93_045421} Both methods have the computational models in which the transition region composed of the objective nanostructures is sandwiched between semi-infinitely continuing electrodes. In the Green's function formalism, the self-energy terms of the electrodes reflect the effect from the semi-infinite electrodes and the perturbed Green's functions of the transition region are computed using the self-energy terms. The conductance of the system is obtained from the Fisher-Lee formula\cite{PhysRevB_23_R6851} by using the perturbed Green's functions and the self-energy terms of the electrodes. On the other hand, in the WFM method, the conductance is expressed as a quantum mechanical scattering problem. The generalized Bloch waves of the electrodes are used to include the contribution of the semi-infinite electrodes in the WFM formula. The conductance is related to the total transmission probability using the transmission coefficients and the group velocity of the Bloch waves. It has been proved that the methods are mathematically equivalent\cite{PhysRevB_86_195406} and the conductances obtained by them should be identical.

Several WFM methods have been proposed so far, and their formulations are slightly different. The computations used to solve the generalized Bloch waves of the electrodes are time consuming and numerically unstable because of the extremely large and small eigenvalues of the Bloch factors.\cite{PhysRevB_67_195315} On the basis of the physical observation that only propagating and slowly decaying evanescent waves of the electrodes contribute to the transmission of electrons, the rapidly decaying evanescent waves can be excluded by introducing a cutoff parameter for the Bloch factor $\lambda_{\text{min}}$ ($|\lambda_{\text{min}}|<1$) in some WFM methods.\cite{PhysRevB_70_195402,PhysRevB_79_205322,JChemPhys_147_074116,arXiv:1709.09324} There have been several discussions on the translational invariance of the transport properties with respect to moving the matching planes in the WFM formalism with the cutoff parameter $\lambda_{\text{min}}$. Krsti\'c {\it et al.} stated that Ando's formulation for the scattering process lacks translational invariance\cite{PhysRevB_44_8017,PhysRevB_66_205319} and, later, Khomyakov {\it et al.} proved that translational invariance is retained when the matching planes are moved.\cite{PhysRevB_70_195402} S\o{}rensen {\it et al.} reported that the accuracy of the transmission probability is degraded when the rapidly varying evanescent waves are excluded.\cite{PhysRevB_79_205322} In addition, the difference in the numerical errors between the transmission and reflection probabilities indicates that the sum of these probabilities does not correspond to the number of channels. They also proposed a method that extends the transition region by inserting a couple of extra layers to improve the accuracy of the transmission probability. However, from the viewpoint of the penetration of quantum particles as well as the proof proposed by Khomyakov {\it et al.}, the improvement of S\o{}rensen {\it et al.} contradicts the fact that the transmission probability does not change when the extra layers are inserted. Therefore, the problems concerning the translational invariance and the sum of the transmission and reflection probabilities are important issues to be resolved for the present WFM methods.

In this paper, we reformulate the WFM formalism because the most of the WFM methods that have been introduced so far include all the propagating and evanescent waves. Then, we explore the origin of the deterioration of the translational invariance presented by S\o{}rensen {\it et al.} It is revealed that the pseudoinverses of the generalized Bloch wave matrices used in the computation of the transmission coefficients degrade the translational invariance and degrade the accuracy of the transmission probability. The error in the transmission probability with respect to the number of extra layers demonstrated in Ref.~\onlinecite{PhysRevB_79_205322} is not due to the WFM methods but rather due to the usage of pseudoinverses. Moreover, we propose a method that can compute the transmission and reflection probabilities without inclusion of the rapidly varying evanescent waves or the extension of the transition region. We find that, in our WFM formalism, the translational invariance is nicely preserved, and the sum of the probabilities exactly corresponds with the number of channels. We also find that the number of iterations for the continued-fraction equations to compute the self-energy terms of the electrodes is closely related to the number of extra layers inserted in the transition region. We demonstrate that the numerical accuracy of our WFM formalism is comparable with that of the nonequilibrium Green's function method even when the rapidly varying evanescent waves are not explicitly computed and the transition region is not extended.

In addition to the improvement of the accuracy in the wave-function matching calculations, we perform practical electron transport calculations of two-dimensional graphene sheets embedded with B--N line defects connected to a pair of semi-infinite graphene electrodes. 
Graphene sheets are known to have only a characteristic band structure at around the Fermi energy $E_{\text{F}}$, the so-called Dirac cone. 
This means that the electrons passing through the junctions are provided only from the Dirac cone, and have a limited range of momenta. 
Through the transport calculations, we present the case that the defect states not matching with the incident-wave modes do not directly contribute to the electron transport, e.g., through resonant transport, but they have indirect influence on the transport properties via hybridization of the Dirac cone and the defect states. 

\section{Formalism}
\label{sec: Formalism}
\subsection{Generalized Bloch waves in electrodes}
\label{sec:Generalized Bloch waves in electrodes}
\begin{figure*}[htb]
\begin{center}
\includegraphics{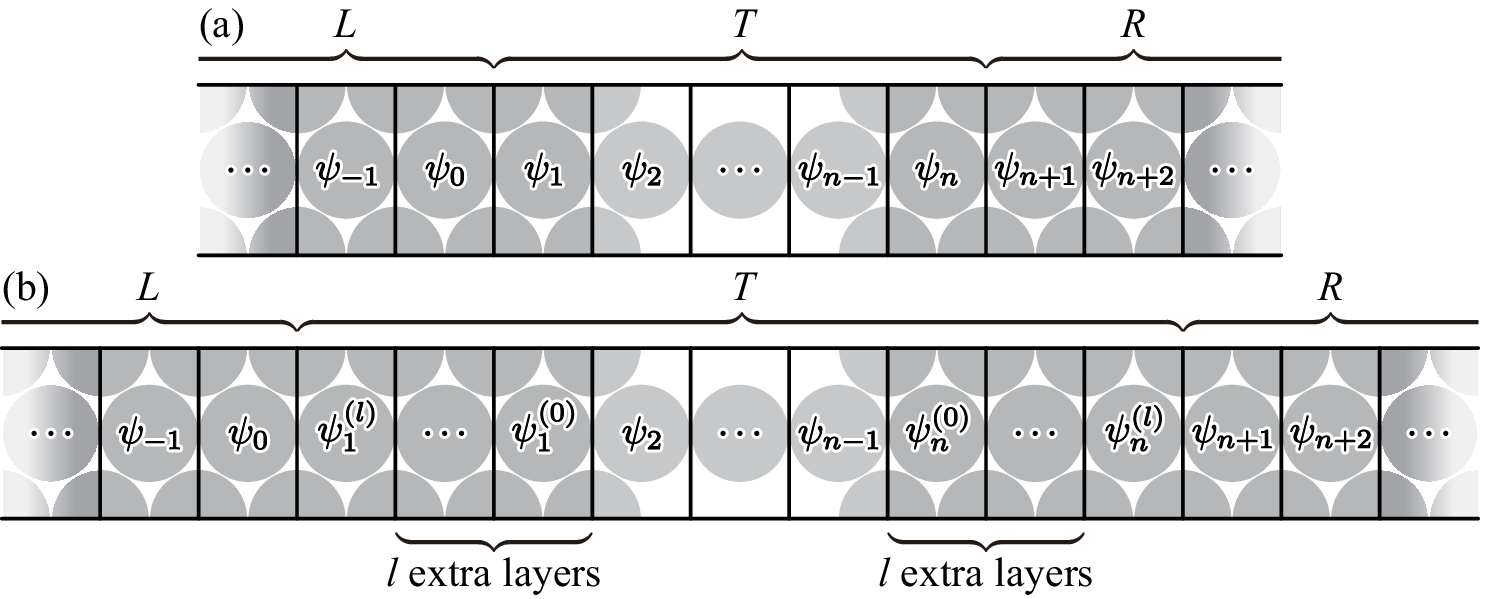}
\caption{Schematic illustration of computational model for WFM formalism. The transition region between the two semi-infinite electrodes represents the objective nanostructures. (a) Model without extra layers in the transition region and (b) model with $l$ extra layers. \label{fig:1}}
\end{center}
\end{figure*}

Let us introduce the WFM formalism for calculating the solution of the Kohn-Sham equation\cite{PhysRev_140_A1133,RevModPhys_71_001253} in a system with a transition region sandwiched between semi-infinitely continuing crystalline electrodes, as shown in Fig.~\ref{fig:1}(a). The solution we wish to calculate is the SWFs specified by particular incident Bloch waves coming from deep inside the left electrode. The SWFs for the Bloch waves coming from the right electrode can be obtained in a similar manner. Since the Kohn-Sham effective potential in a crystalline electrode is periodic, the wave functions $\vec{\phi}_{(j)}$ in the $j$th unit cell of the electrode satisfy the generalized Bloch condition,
\begin{equation}
\label{eqn:01-01}
\vec{\phi}_{(j)}=\lambda^{j-i}\vec{\phi}_{(i)},
\end{equation}
where $i$ and $j$ are the indices of the unit cell of the crystal. Making use of the generalized Bloch condition, the wave functions in the left electrode $\vec{\phi}_{(i)}$ are obtained by solving the following quadratic eigenvalue problem for the given energy $E$;\cite{PhysRevB_44_8017}
\begin{equation}
\label{eqn:01-02}
\overline{\mat{H}}^\dag_{L,L} \vec{\phi}_{(i)} + \lambda \overline{\mat{H}}_{L} \vec{\phi}_{(i)}+ \lambda^2 \overline{\mat{H}}_{L,L} \vec{\phi}_{(i)} = 0.
\end{equation}
Here, $\overline{\mat{H}}_{L}$ is the $m_L \times m_L$ full rank matrix for the Hamiltonian of the periodic unit cell with $m_L$ being the number of real-space grids or bases in a unit cell and $\overline{\mat{H}}_L \equiv E\mat{S}_L-\mat{H}_L$ with $\mat{S}_L$ being the overlap matrix of the bases. If one uses a real-space grid method and norm-conserving pseudopotentials,\cite{PhysRevB_43_001993} $\mat{S}_L$ is an identity matrix. $\overline{\mat{H}}_{L,L}$ is the $m_L \times m_L$ off-diagonal block matrix of the infinite Hamiltonian of the electrode representing the coupling between two neighboring unit cells. Because $\overline{\mat{H}}_{L,L}$ is not a full rank matrix in some cases, the rank of $\overline{\mat{H}}_{L,L}$ is represented by the other variable $m_{LL}$. In the case of $m_L > m_{LL}$, we assume that the $\overline{\mat{H}}_{LL}$ is a zero matrix except that the lower left $m_{LL} \times m_{LL}$ block-matrix element is $\overline{\mat{h}}_{L}$,
\begin{eqnarray}
\label{eqn:01-03}
\overline{\mat{H}}_{L,L}=\left(
\begin{array}{ccc}
0 & \cdots & 0 \\
\vdots & \ddots & \vdots \\
\overline{\mat{h}}^L & \cdots & 0 \\
\end{array}
\right).
\end{eqnarray}
This assumption does not affect generality because $\overline{\mat{H}}_{LL}$ can be described using Eq.~(\ref{eqn:01-03}) by the unitary transformation.\cite{PhysRevE_90_013306,PhysRevE_95_033309} The solution of Eq.~(\ref{eqn:01-02}) is obtained by solving the following eigenvalue problem:
\begin{eqnarray}
\label{eqn:01-03a}
\left(
\begin{array}{cc}
\overline{\mat{H}_L} & \overline{\mat{H}}_{L,L}^\dag \\
\mat{I} & \mat{0}
\end{array}
\right)
\left(
\begin{array}{c}
\vec{\phi}_{(i)} \\
\vec{\phi}_{(i-1)} \\
\end{array}
\right)
=
\lambda
\left(
\begin{array}{cc}
-\overline{\mat{H}}_{L,L} & \mat{0} \\
\mat{0} & \mat{I}
\end{array}
\right)
\left(
\begin{array}{c}
\vec{\phi}_{(i)} \\
\vec{\phi}_{(i-1)} \\
\end{array}
\right).
\end{eqnarray}
Ando proposed a method of solving Eq.~(\ref{eqn:01-03a}) as a standard eigenvalue problem, computational cost of which is $O(m_L^3)$.\cite{PhysRevB_44_8017}

In 2008, S\o{}rensen {\it et al.} reported a procedure that obtains selected interior eigenpairs of large-scale general complex eigenproblems, $\lambda_{\text{min}}<|\lambda|<1/\lambda_{\text{min}}$, by using an iterative Krylov subspace technique, where $\lambda_{\text{min}}$ is the cutoff parameter of the Bloch factor.\cite{PhysRevB_77_155301} Laux proposed a practical approach which resolves only eigenpairs within a contour defined in the complex $\lambda$ plane.\cite{PhysRevB_86_075103} Later, this approach of utilizing contour integrations was applied to the real-space grid schemes with some improvements in the Sakurai-Sugiura method.\cite{SC17,arXiv:1709.09324,JComputApplMath_159_119} Although they do not obtain all eigenpairs required to treat a semi-infinite system, these approaches are valid within the assumption that only the propagating and slowly decreasing evanescent waves contribute to the transport properties.\cite{PhysRevB_70_195402,PhysRevB_79_205322}

The alternative approach is to use the Green's function of the isolated Hamiltonian of the electrode $\mat{\Theta}_L [\equiv (\overline{\mat{H}}_{L})^{-1}]$ proposed by Fujimoto and Hirose.\cite{PhysRevB_67_195315} In this scheme, one assumes that $m_L > 2m_{LL}$. If $m_L < 2m_{LL}$, several unit cells of the crystal are included in $\overline{\mat{H}}_{L}$ to satisfy $m_L > 2m_{LL}$. The generalized Bloch waves are obtained by solving the following generalized eigenvalue problem
\begin{eqnarray}
\label{eqn:01-05}
\left(
\begin{array}{cc}
- \bm{\thetaup}_{L}(\xi_3,\xi_1) \overline{\mat{h}}^{L \dag} & - \bm{\thetaup}_{L}(\xi_3,\xi_3) \overline{\mat{h}}^L \\
\mat{0} & \mat{I}
\end{array}
\right)
\left(
\begin{array}{c}
\vec{\phi}_{(i-1)}(\xi_3) \\
\vec{\phi}_{(i+1)}(\xi_1)
\end{array}
\right)= \nonumber \\
\lambda
\left(
\begin{array}{cc}
\mat{I} & \mat{0} \\
- \bm{\thetaup}_{L}(\xi_1,\xi_1) \overline{\mat{h}}^{L \dag} & - \bm{\thetaup}_{L}(\xi_1,\xi_3) \overline{\mat{h}}^L \\
\end{array}
\right)
\left(
\begin{array}{c}
\vec{\phi}_{(i-1)}(\xi_3) \\
\vec{\phi}_{(i+1)}(\xi_1)
\end{array}
\right),
\end{eqnarray}
where
\begin{eqnarray}
\label{eqn:01-04}
\mat{\Theta}_L \equiv (\overline{\mat{H}}_{L})^{-1}= \left(
\begin{array}{ccc}
\bm{\thetaup}_{L}(\xi_1,\xi_1) & \bm{\thetaup}_{L}(\xi_1,\xi_2) & \bm{\thetaup}_{L}(\xi_1,\xi_3) \\
\bm{\thetaup}_{L}(\xi_2,\xi_1) & \bm{\thetaup}_{L}(\xi_2,\xi_2) & \bm{\thetaup}_{L}(\xi_2,\xi_3) \\
\bm{\thetaup}_{L}(\xi_3,\xi_1) & \bm{\thetaup}_{L}(\xi_3,\xi_2) & \bm{\thetaup}_{L}(\xi_3,\xi_3)
\end{array}
\right).
\end{eqnarray}
Here, multiple grids or bases are bunched together in $\xi_i$. $\xi_1$ $(\xi_3)$ includes $m_{LL}$ grids or bases having interactions with the left (right) neighboring unit cell $\{\chi_{1,j}\}$ ($\{\chi_{3,j}\}$) and $\xi_2$ gathers $m_L-2m_{LL}$ grids or bases not having the interaction with the neighboring cells $\{\chi_{2,j}\}$. For example,
\begin{equation}
\xi_1 = (\chi_{1,1}, \cdots, \chi_{1,m_{LL}}).
\end{equation}
Thus, $\bm{\thetaup}_{L}(\xi_1,\xi_1)$ is an $m_{LL} \times m_{LL}$ matrix expressed as $\bm{\thetaup}_{L}(\xi_1,\xi_1)=\langle \xi_1|\mat{\Theta}_L| \xi_1 \rangle$. The computational cost of this scheme is $O(m_Lm_{LL})$ to set up Eq.~(\ref{eqn:01-05}), when $\overline{\mat{H}}_{L}$ is sparse and an iterative method is used. Although this scheme computes the eigenpairs of Eq.~(\ref{eqn:01-05}) within the specific interval of $\lambda_{\text{min}}<|\lambda|<1/\lambda_{\text{min}}$, truly semi-infinite systems can be treated by solving the continued-fraction equations introduced in Sec.~\ref{sec:Moving matching plane of wave function matching formula}.

The generalized Bloch waves are evenly divided into two groups, and the number of waves in a group is $m_{LL}$. The eigenpairs $\lambda^{-}_k$ and $\vec{\phi}^{-}_{(i),k}$ ($\lambda^{+}_k$ and $\vec{\phi}^{+}_{(i),k}$) represent the group of the left (right) decreasing evanescent waves $|\lambda_k|>0$ ($|\lambda_k|<0$) and left (right) propagating waves $|\lambda_k|=1$. We also introduce the Bloch matrix, which relates the generalized Bloch waves with those in the neighboring cells,
\begin{equation}
\label{eqn:01-06}
\mat{B}^{\pm}_L=\mat{\Phi}_{L(i+1)}^{\pm}(\mat{\Phi}_{L(i)}^{\pm})^{-1}=\mat{\Phi}_{L(i)}^{\pm} \mat{\Lambda}_L^{\pm}(\mat{\Phi}_{L(i)}^{\pm})^{-1},
\end{equation}
where $\mat{\Lambda}^\pm_{L}=\mbox{diag}[\lambda^\pm_1,\cdots,\lambda^\pm_{m_{LL}}]$ and $\mat{\Phi}^{\pm}_{L(i)}=[\vec{\phi}^{\pm}_{(i),1}, \cdots, \vec{\phi}^{\pm}_{(i),m_{LL}}]$. Equation~(\ref{eqn:01-06}) leads to the relation,
\begin{equation}
\label{eqn:01-07}
\mat{\Phi}_{L(j)}^{\pm}=\left( \mat{B}^{\pm}_L \right)^{j-i} \mat{\Phi}_{L(i)}^{\pm}=\mat{\Phi}_{L(i)}^{\pm} \left (\mat{\Lambda}_L^{\pm} \right)^{j-i}.
\end{equation}
Hereafter, the index $(i)$ for the unit cell is left out because it only affects the trivial Bloch factor in Eq.~(\ref{eqn:01-07}). For the right electrode, the same quantities are defined with $L \rightarrow R$.

\subsection{Expression using transmission and reflection coefficients}
In the WFM formalism, the SWFs in the right electrode region are expressed as a linear combination of generalized Bloch waves, 
\begin{equation}
\label{eqn:02-01}
\vec{\psi}_{n+1,k} = {\mat{\Phi}}_{R}^{+} \vec{t}_{k},
\end{equation}
where $\vec{t}_{k}$ is an $m_R$-dimensional vector containing the transmission coefficients and $k$ is the index of the incident waves. The SWFs in the left electrode region are defined as 
\begin{eqnarray}
\label{eqn:02-02}
\vec{\psi}_{0,k} &=& {\mat{\Phi}}_{L}^{-} \vec{r}_{k} + \vec{\phi}^+_{L,k} \nonumber \\
&=& \vec{\psi}^{ref}_{0,k} + \vec{\psi}^{in}_{0,k},
\end{eqnarray}
where
\begin{eqnarray}
\label{eqn:02-03}
\vec{\psi}^{ref}_{0,k} &=& {\mat{\Phi}}_{L}^{-} \vec{r}_{k}, \\
\label{eqn:02-04}
\vec{\psi}^{in}_{0,k} &=& \vec{\phi}^+_{L,k}
\end{eqnarray}
with $\vec{r}_{k}$ being an $m_L$-dimensional vector containing the reflection coefficients. Here, $\vec{\psi}^{ref}_{0,k}$ and $\vec{\psi}^{in}_{0,k}$ represent the reflection and incident waves, respectively. In addition, the transmission and reflection coefficients are given by
\begin{eqnarray}
\label{eqn:02-05}
\vec{t}_{k} &=& ({\mat{\Phi}}_{R}^{+})^{-1} \vec{\psi}_{n+1,k}, \\
\label{eqn:02-06}
\vec{r}_{k} &=& ({\mat{\Phi}}_{L}^{-})^{-1} \vec{\psi}^{ref}_{0,k}.
\end{eqnarray}

\subsection{Exclusion of rapidly varying evanescent waves}
\label{sec:Excluding evanescent modes}
The coefficients for the extremely fast decaying evanescent waves are very small when the matching planes are set far from the scatterers. S\o{}rensen {\it et al.} split $\mat{\Phi}^\pm $ into propagating and moderately decaying evanescent waves $\tilde{\mat{\Phi}}^\pm $ and rapidly decaying evanescent waves $\mathring{\mat{\Phi}}^\pm $\cite{PhysRevB_79_205322}
\begin{equation}
\label{eqn:03-01}
\mat{\Phi}^\pm = [\tilde{\mat{\Phi}}^\pm,\mathring{\mat{\Phi}}^\pm].
\end{equation}
The SWFs at the first layers of the right and left electrode regions are rewritten as
\begin{eqnarray}
\label{eqn:03-02}
\vec{\psi}_{n+1,k}^+ = \mat{\Phi}^+ \vec{a}_{n+1,k}^+ = [\tilde{\mat{\Phi}}^+,\mathring{\mat{\Phi}}^+]
\left(
\begin{array}{c}
\vec{\tilde{a}}_{n+1,k}^+ \\
\vec{\mathring{a}}_{n+1,k}^+
\end{array}
\right),
\end{eqnarray}
and
\begin{eqnarray}
\label{eqn:03-03}
\vec{\psi}_{0,k}^{ref,-} = \mat{\Phi}^- \vec{a}_{0,k}^{ref,-} = [\tilde{\mat{\Phi}}^-,\mathring{\mat{\Phi}}^-]
\left(
\begin{array}{c}
\vec{\tilde{a}}_{0,k}^{ref,-} \\
\vec{\mathring{a}}_{0,k}^{ref,-}
\end{array}
\right),
\end{eqnarray}
respectively, where $\vec{a}_{i,k}^\pm=[\tilde{\vec{a}}^{\pm T}_{i,k},\vec{\mathring{a}}^{\pm T}_{i,k}]^T$ are vectors that contain the expansion coefficients of the generalized Bloch waves.

It is computationally demanding and numerically unstable to obtain all eigenpairs of Eq.~(\ref{eqn:01-03a}) or (\ref{eqn:01-05}). To retain numerical accuracy without $\mathring{\mat{\Phi}}^\pm$, S\o rensen {\it et al.} proposed a method that inserts extra layers in the transition region, as shown in Fig.~\ref{fig:1}(b). Making use of the relation $\mat{B}^{\pm}=\mat{\Phi}^{\pm}\mat{\Lambda}^{\pm}(\mat{\Phi}^{\pm})^{-1}$, the SWFs in the electrode regions are expressed as 
\begin{eqnarray}
\label{eqn:03-04}
\vec{\psi}_{n+1,k}^{+}=(\mat{B}_{R}^{+})^{l}\vec{\psi}_{n,k}^{(0),+}=\left(
\begin{array}{c}
\tilde{\mat{\Phi}}_{R}^{+}, \mathring{\mat{\Phi}}_{R}^{+}
\end{array}
\right)\left(
\begin{array}{c}
(\tilde{\mat{\Lambda}}_{R}^{+})^{l}\vec{\tilde{a}}_{n,k}^{(0),+} \\
(\mathring{\mat{\Lambda}}_{R}^{+})^{l}\vec{\mathring{a}}_{n,k}^{(0),+}
\end{array}
\right), \\
\label{eqn:03-05}
\vec{\psi}_{0,k}^{ref,-}=(\mat{B}_{L}^{-})^{-l}\vec{\psi}_{1,k}^{ref,(0),-}=\left(
\begin{array}{c}
\tilde{\mat{\Phi}}_{L}^{-}, \mathring{\mat{\Phi}}_{L}^{-}
\end{array}
\right)\left(
\begin{array}{c}
(\tilde{\mat{\Lambda}}_{L}^{-})^{-l}\vec{\tilde{a}}_{1,k}^{ref,(0),-} \\
(\mathring{\mat{\Lambda}}_{L}^{-})^{-l}\vec{\mathring{a}}_{1,k}^{ref,(0),-}
\end{array}
\right).
\end{eqnarray}
From Eqs.~(\ref{eqn:02-05}) and (\ref{eqn:02-06}), we have
\begin{eqnarray}
\label{eqn:03-06}
\vec{t}_k &=& \left(
\begin{array}{c}
\tilde{\vec{t}}_k \\
\mathring{\vec{t}}_k
\end{array}
\right) = \left(
\begin{array}{c}
\tilde{\mat{\Phi}}_{R}^{+}, \mathring{\mat{\Phi}}_{R}^{+}
\end{array}
\right)^{-1}\vec{\psi}_{n+1,k}^{+} = \left(
\begin{array}{c}
(\tilde{\mat{\Lambda}}_{R}^{+})^{l}\vec{\tilde{a}}_{n,k}^{(0),+} \\
(\mathring{\mat{\Lambda}}_{R}^{+})^{l}\vec{\mathring{a}}_{n,k}^{(0),+}
\end{array}
\right), \\
\label{eqn:03-07}
\vec{r}_k &=& \left(
\begin{array}{c}
\tilde{\vec{r}}_k \\
\mathring{\vec{r}}_k
\end{array}
\right)
= \left(
\begin{array}{c}
\tilde{\mat{\Phi}}_{L}^{-}, \mathring{\mat{\Phi}}_{L}^{-}
\end{array}
\right)^{-1}\vec{\psi}_{0,k}^{ref,-}
= \left(
\begin{array}{c}
(\tilde{\mat{\Lambda}}_{L}^{-})^{-l}\vec{\tilde{a}}_{1,k}^{ref,(0),-} \\
(\mathring{\mat{\Lambda}}_{L}^{-})^{-l}\vec{\mathring{a}}_{1,k}^{ref,(0),-}
\end{array}
\right).
\end{eqnarray}
The coefficients of the rapidly decaying evanescent waves vanish when a sufficient number of extra layers are inserted, because the Bloch factors of the rapidly decaying evanescent waves, $1/|\lambda^{-}_L| (< \lambda_{\text{min}})$ and $|\lambda^{+}_R| (< \lambda_{\text{min}})$, are much smaller than 1.
\begin{eqnarray}
\label{eqn:03-08}
(\mathring{\mat{\Lambda}}_{R}^{+})^{l}\vec{\mathring{a}}_{n,k}^{(0),+} \approx \mat{0}, \\
\label{eqn:03-09}
(\mathring{\mat{\Lambda}}_{L}^{-})^{-l}\vec{\mathring{a}}_{1,k}^{ref,(0),-} \approx \mat{0}.
\end{eqnarray}
Consequently, the SWFs at the first layers of the electrode regions are rewritten as
\begin{equation}
\label{eqn:03-10}
\vec{\psi}_{n+1,k} = \vec{\psi}_{n+1,k}^{+} = \left(
\begin{array}{c}
\tilde{\mat{\Phi}}_{R}^{+}, \mathring{\mat{\Phi}}_{R}^{+}
\end{array}
\right)\left(
\begin{array}{c}
(\tilde{\mat{\Lambda}}_{R}^{+})^{l}\vec{\tilde{a}}_{n,k}^{(0),+} \\
(\mathring{\mat{\Lambda}}_{R}^{+})^{l}\vec{\mathring{a}}_{n,k}^{(0),+}
\end{array}
\right) \approx \tilde{\mat{\Phi}}_{R}^{+} (\tilde{\mat{\Lambda}}_{R}^{+})^{l}\vec{\tilde{a}}_{n,k}^{(0),+},
\end{equation}
\begin{eqnarray}
\label{eqn:03-11}
\left\{
\begin{array}{l}
\vec{\psi}_{0,k} = \vec{\psi}_{0,k}^{ref} + \vec{\psi}_{0,k}^{in}, \\
\vec{\psi}_{0,k}^{ref} = \vec{\psi}_{0,k}^{ref,-} = \left(
\begin{array}{c}
\tilde{\mat{\Phi}}_{L}^{-}, \mathring{\mat{\Phi}}_{L}^{-}
\end{array}
\right)\left(
\begin{array}{c}
(\tilde{\mat{\Lambda}}_{L}^{-})^{-l}\vec{\tilde{a}}_{1,k}^{ref,(0),-} \\
(\mathring{\mat{\Lambda}}_{L}^{-})^{-l}\vec{\mathring{a}}_{1,k}^{ref,(0),-}
\end{array}
\right)\approx \tilde{\mat{\Phi}}_{L}^{-} (\tilde{\mat{\Lambda}}_{L}^{-})^{-l}\vec{\tilde{a}}_{1,k}^{ref,(0),-} \\
\vec{\psi}_{0,k}^{in} = (\lambda_{L,k}^+)^{-l} \vec{\phi}^+_{L,k}
\end{array}
\right. .
\end{eqnarray}
The SWFs at the second layers in the electrode regions are expressed in the same manner:
\begin{equation}
\label{eqn:03-12}
\vec{\psi}_{n+2,k} = \vec{\psi}_{n+2,k}^{+} =
\left(
\begin{array}{c}
\tilde{\mat{\Phi}}_{R}^{+}\tilde{\mat{\Lambda}}_{R}^{+}, \mathring{\mat{\Phi}}_{R}^{+}\mathring{\mat{\Lambda}}_{R}^{+}
\end{array}
\right)\left(
\begin{array}{c}
(\tilde{\mat{\Lambda}}_{R}^{+})^{l}\vec{\tilde{a}}_{n,k}^{(0),+} \\
(\mathring{\mat{\Lambda}}_{R}^{+})^{l}\vec{\mathring{a}}_{n,k}^{(0),+}
\end{array}
\right)
\approx \tilde{\mat{\Phi}}_{R}^{+} (\tilde{\mat{\Lambda}}_{R}^{+})^{l+1}\vec{\tilde{a}}_{n,k}^{(0),+},
\end{equation}
\begin{eqnarray}
\label{eqn:03-13}
\left\{
\begin{array}{l}
\vec{\psi}_{-1,k} = \vec{\psi}_{-1,k}^{ref} + \vec{\psi}_{-1,k}^{in}, \\
\vec{\psi}_{-1,k}^{ref} = \vec{\psi}_{-1,k}^{ref,-} =
\left(
\begin{array}{c}
\tilde{\mat{\Phi}}_{L}^{-}(\tilde{\mat{\Lambda}}_{L}^{-})^{-1}, \mathring{\mat{\Phi}}_{L}^{-}(\mathring{\mat{\Lambda}}_{L}^{-})^{-1}
\end{array}
\right)\left(
\begin{array}{c}
(\tilde{\mat{\Lambda}}_{L}^{-})^{-l}\vec{\tilde{a}}_{1,k}^{ref,(0),-} \\
(\mathring{\mat{\Lambda}}_{L}^{-})^{-l}\vec{\mathring{a}}_{1,k}^{ref,(0),-}
\end{array}
\right)
\approx \tilde{\mat{\Phi}}_{L}^{-} (\tilde{\mat{\Lambda}}_{L}^{-})^{-l-1}\vec{\tilde{a}}_{1,k}^{ref,(0),-} \\
\vec{\psi}_{-1,k}^{in} = (\lambda_{L,k}^+)^{-l-1} \vec{\phi}^+_{L,k}
\end{array}
\right. .
\end{eqnarray}
Hereafter, we will leave out the symbols $+$ and $-$ because it is obvious which generalized Bloch waves are included in the SWFs when the index of the layer is indicated. In addition, the index of the incident waves $k$ will be also omitted for notational simplicity.

\subsection{Boundary condition}
\label{sec:Boundary condition}
Table I summarizes the expressions for the electrode regions and transition region in the WFM methods proposed so far. The electrode regions are the layers where the transmitted and reflected waves are defined by a linear combination of generalized Bloch waves as in Eqs.~(\ref{eqn:02-01}) and (\ref{eqn:02-03}), respectively. The transition region corresponds to the layers in which the values on real-space grids or coefficients for the atomic bases of the SWFs are calculated using the WFM procedure. In the previous WFM methods, the values or coefficients at the edge layers of the transition region are defined on real-space grids or expanded using an atomic basis set, whereas they are given by a linear combination of generalized Bloch waves as a boundary condition in the electrode regions (e.g., Eqs.~(\ref{eqn:02-01}) and (\ref{eqn:02-03})). As long as the complete set of generalized Bloch waves is employed, the SWFs determined in the electrode and transition regions are identical. However, numerical errors will occur when the rapidly varying evanescent waves are excluded by the cutoff parameter of the Bloch factor $\lambda_{\text{min}}$. In our procedure, to keep numerical rigorousness, the overlap of the layers between the electrode and transition regions is eliminated.

\begin{table}
\centering
\caption{Electrode and transition regions for the WFM methods proposed so far. \label{tbl:1}}
\begin{tabular}{cccc}
\hline \hline 
& left electrode region & transition region & right electrode region \\ \hline
Fujimoto \& Hirose\cite{PhysRevB_67_195315} & $\cdots, -1, 0$ & $0, 1, \cdots, n, n+1$ & $n+1, n+2, \cdots$ \\
Khomyakov {\it et al.}\cite{PhysRevB_70_195402} & $\cdots, -1, 0$ & $0, 1, \cdots, n, n+1$ & $n+1, n+2, \cdots$ \\
S\o{}rensen {\it et al.}\cite{PhysRevB_79_205322} & $\cdots, 0, 1$ & $1, 2, \cdots, n-1, n$ & $n, n+1, \cdots$ \\
Present work & $\cdots, -1, 0$ & $1, 2, \cdots, n-1, n$ & $n+1, n+2, \cdots$ \\
\hline \hline 
\end{tabular}
\end{table}

\subsection{Wave function matching formula}
\label{sec:Wave function matching formula}
Here let us introduce the WFM formula\cite{PhysRevB_67_195315} for the case that the rank of $\overline{\mat{H}}_{L}(\overline{\mat{H}}_{R})$, $m_L(m_R)$, and the rank of $\overline{\mat{H}}_{L,L}(\overline{\mat{H}}_{R,R})$, $m_{LL}(m_{RR})$, are equal. The case that they are not equal is formulated in Appendix~\ref{sec:Wave function matching for general case}. By letting $\widehat{\mat{H}}^{(0)}$ be the $m_T$-dimensional Hamiltonian for the computational model without any extra layers [Fig.~\ref{fig:1}(a)], the Kohn-Sham equation for the computational model with $l$ extra layers [Fig.~\ref{fig:1}(b)] is expressed as
\begin{equation}
\label{eqn:04-01}
E\widehat{\mat{S}}^{(l)}-\widehat{\mat{H}}^{(l)}
\left(
\begin{array}{c}
\vec{\psi}_{1}^{(l)} \\
\vdots \\
\vdots \\
\vdots \\
\vec{\psi}_{n}^{(l)}
\end{array}
\right)
=
\left(
\begin{array}{c}
-\overline{\mat{H}}_{L,L}^\dag\vec{\psi}_{0} \\
0 \\
\vdots \\
0 \\
-\overline{\mat{H}}_{R,R}\vec{\psi}_{n+1}
\end{array}
\right)
\end{equation}
using the [$m_T+l(m_L+m_R)$]-dimensional Hamiltonian $\widehat{\mat{H}}^{(l)}$ and the overlap matrix $\widehat{\mat{S}}^{(l)}$.
Multiplying the Green's function $\widehat{\mat{G}}^{(l)}[=(E\widehat{\mat{S}}^{(l)}-\widehat{\mat{H}}^{(l)})^{-1}]$ from the left hand side yields
\begin{equation}
\label{eqn:04-02}
\left(
\begin{array}{c}
\vec{\psi}_{1}^{(l)} \\
\vdots \\
\vdots \\
\vdots \\
\vec{\psi}_{n}^{(l)}
\end{array}
\right)
=
\widehat{\mat{G}}^{(l)}\left(
\begin{array}{c}
-\overline{\mat{H}}_{L,L}^\dag\vec{\psi}_{0} \\
0 \\
\vdots \\
0 \\
-\overline{\mat{H}}_{R,R}\vec{\psi}_{n+1}
\end{array}
\right),
\end{equation}
where 
\begin{eqnarray}
\label{eqn:04-03}
\widehat{\mat{G}}^{(l)} = \left(
\begin{array}{ccccccccc}
\mat{G}^{(l)}_{1-l,1-l} & \cdots & \mat{G}^{(l)}_{1-l,0} & \mat{G}^{(l)}_{1-l,1} & \cdots & \mat{G}^{(l)}_{1-l,n} & \mat{G}^{(l)}_{1-l,n+1} & \cdots & \mat{G}^{(l)}_{1-l,n+l} \\
\vdots & \ddots & \vdots & \vdots & & \vdots & \vdots & & \vdots  \\
\mat{G}^{(l)}_{0,1-l} & \cdots & \mat{G}^{(l)}_{0,0} & \mat{G}^{(l)}_{0,1} & \cdots & \mat{G}^{(l)}_{0,n} & \mat{G}^{(l)}_{0,n+1} & \cdots & \mat{G}^{(l)}_{0,n+l} \\
\mat{G}^{(l)}_{1,1-l} & \cdots & \mat{G}^{(l)}_{1,0} & \mat{G}^{(l)}_{1,1} & \cdots & \mat{G}^{(l)}_{1,n} & \mat{G}^{(l)}_{1,n+1} & \cdots & \mat{G}^{(l)}_{1,n+l} \\
\vdots & & \vdots & \vdots & \ddots & \vdots & \vdots & & \vdots  \\
\mat{G}^{(l)}_{n,1-l} & \cdots & \mat{G}^{(l)}_{n,0} & \mat{G}^{(l)}_{n,1} & \cdots & \mat{G}^{(l)}_{n,n} & \mat{G}^{(l)}_{n,n+1} & \cdots & \mat{G}^{(l)}_{n,n+l} \\
\mat{G}^{(l)}_{n+1,1-l} & \cdots & \mat{G}^{(l)}_{n+1,0} & \mat{G}^{(l)}_{n+1,1} & \cdots & \mat{G}^{(l)}_{n+1,n} & \mat{G}^{(l)}_{n+1,n+1} & \cdots & \mat{G}^{(l)}_{n+1,n+l} \\
\vdots & & \vdots & \vdots & & \vdots & \vdots & \ddots & \vdots  \\
\mat{G}^{(l)}_{n+l,1-l} & \cdots & \mat{G}^{(l)}_{n+l,0} & \mat{G}^{(l)}_{n+l,1} & \cdots & \mat{G}^{(l)}_{n+l,n} & \mat{G}^{(l)}_{n+l,n+1} & \cdots & \mat{G}^{(l)}_{n+l,n+l} \\
\end{array}
\right)
\end{eqnarray}
with $\mat{G}^{(l)}_{i,j}$ being the $(i, j)$th block-matrix element of the Green's function of the isolated transition region $\widehat{\mat{G}}^{(l)}$. From the first and last block rows of Eq.~(\ref{eqn:04-02}), $\vec{\psi}_{0}$ and $\vec{\psi}_{n+1}$ are related to $\vec{\psi}_{1}^{(l)}$ and $\vec{\psi}_{n}^{(l)}$ as follows:
\begin{eqnarray}
\label{eqn:04-04}
\vec{\psi}_{1}^{(l)} &=& - \mat{G}^{(l)}_{1-l,1-l}\overline{\mat{H}}_{L,L}^\dag\vec{\psi}_{0} - \mat{G}^{(l)}_{1-l,n+l}\overline{\mat{H}}_{R,R}\vec{\psi}_{n+1}, \\
\label{eqn:04-05}
\vec{\psi}_{n}^{(l)} &=& - \mat{G}^{(l)}_{n+l,1-l}\overline{\mat{H}}_{L,L}^\dag\vec{\psi}_{0} - \mat{G}^{(l)}_{n+l,n+l}\overline{\mat{H}}_{R,R}\vec{\psi}_{n+1}.
\end{eqnarray}
Substituting the relation of Eq.~(\ref{eqn:02-02}) into Eqs.~(\ref{eqn:04-04}) and (\ref{eqn:04-05}), we obtain
\begin{eqnarray}
\label{eqn:04-06}
\vec{\psi}_{1}^{ref,(l)} + \vec{\psi}_{1}^{in,(l)} &=& - \mat{G}^{(l)}_{1-l,1-l}\overline{\mat{H}}_{L,L}^\dag(\vec{\psi}_{0}^{ref} + \vec{\psi}_{0}^{in}) - \mat{G}^{(l)}_{1-l,n+l}\overline{\mat{H}}_{R,R}\vec{\psi}_{n+1}, \\
\label{eqn:04-07}
\vec{\psi}_{n}^{(l)} &=& - \mat{G}^{(l)}_{n+l,1-l}\overline{\mat{H}}_{L,L}^\dag(\vec{\psi}_{0}^{ref} + \vec{\psi}_{0}^{in}) - \mat{G}^{(l)}_{n+l,n+l}\overline{\mat{H}}_{R,R}\vec{\psi}_{n+1}.
\end{eqnarray}
Rewriting Eqs.~(\ref{eqn:04-06}) and (\ref{eqn:04-07}) in terms of the SWFs in the electrode regions $\vec{\psi}_{0}^{ref}$ and $\vec{\psi}_{n+1}$, we arrive at the following WFM formula,
\begin{eqnarray}
\label{eqn:04-08}
\left(
\begin{array}{cc}
- \mat{G}^{(l)}_{1-l,1-l}\overline{\mat{H}}_{L,L}^\dag - \vec{\psi}_{1}^{ref,(l)} \left( \vec{\psi}_{0}^{ref} \right)^{-1} & - \mat{G}^{(l)}_{1-l,n+l}\overline{\mat{H}}_{R,R} \\
- \mat{G}^{(l)}_{n+l,1-l}\overline{\mat{H}}_{L,L}^\dag & - \mat{G}^{(l)}_{n+l,n+l}\overline{\mat{H}}_{R,R} - \vec{\psi}_{n}^{(l)} \left( \vec{\psi}_{n+1} \right)^{-1}
\end{array}
\right)
\left(
\begin{array}{c}
\vec{\psi}_{0}^{ref} \\
\vec{\psi}_{n+1}
\end{array}
\right) \nonumber \\
=
\left(
\begin{array}{c}
\mat{G}^{(l)}_{1-l,1-l}\overline{\mat{H}}_{L,L}^\dag \vec{\psi}_{0}^{in} + \vec{\psi}_{1}^{in,(l)}\\
\mat{G}^{(l)}_{n+l,1-l}\overline{\mat{H}}_{L,L}^\dag \vec{\psi}_{0}^{in}
\end{array}
\right).
\end{eqnarray}
Here, $\vec{\psi}_{0}^{in}$ is determined by Eq.~(\ref{eqn:03-11}) and $\vec{\psi}_{1}^{in,(l)}=\lambda_{L,k}^+\vec{\psi}_0^{in}$. In Eq.~(\ref{eqn:04-08}), the left (right) matching plane is set between $\vec{\psi}_{0}^{ref}$ and $\vec{\psi}_{1}^{ref,(l)}$ ($\vec{\psi}_{n}^{(l)}$ and $\vec{\psi}_{n+1}$).

To solve the simultaneous equations of Eq.~(\ref{eqn:04-08}), we need the ratios of the SWFs between the neighboring layers on both sides of the matching planes $\vec{\psi}_{1}^{ref,(l)} ( \vec{\psi}_{0}^{ref} )^{-1}$ and $\vec{\psi}_{n}^{(l)} \left( \vec{\psi}_{n+1} \right)^{-1}$. From the boundary conditions in the electrode regions, i.e., Eqs.~(\ref{eqn:03-10}), (\ref{eqn:03-11}), (\ref{eqn:03-12}), and (\ref{eqn:03-13}), the ratios in the right and left electrode regions are given by
\begin{eqnarray}
\label{eqn:04-09}
\vec{\psi}_{n+1}(\vec{\psi}_{n+2})^{-1} &=&
\tilde{\mat{\Phi}}_{R}^{+}
\left( \tilde{\mat{\Lambda}}_{R}^{+}\tilde{\mat{\Phi}}_{R}^{+}\right)^{-1},
\end{eqnarray}
and
\begin{eqnarray}
\label{eqn:04-10}
\vec{\psi}_{0}^{ref}(\vec{\psi}_{-1}^{ref})^{-1} &=& \tilde{\mat{\Phi}}_{L}^{-}
\left( (\tilde{\mat{\Lambda}}_{L}^{-})^{-1}\tilde{\mat{\Phi}}_{L}^{-}\right)^{-1},
\end{eqnarray}
respectively. Here, $\tilde{\mat{\Phi}}_{R}^{+}$ and $\tilde{\mat{\Phi}}_{L}^{-}$ contain only the propagating and slowly varying evanescent waves. Thus, $( \tilde{\mat{\Lambda}}_{R}^{+}\tilde{\mat{\Phi}}_{R}^{+})^{-1}$ and $( (\tilde{\mat{\Lambda}}_{L}^{-})^{-1}\tilde{\mat{\Phi}}_{L}^{-})^{-1}$ are determined by the pseudoinverses. In the left electrode region, the SWF satisfies
\begin{equation}
\label{eqn:04-11}
\overline{\mat{H}}^\dag_{L,L} \vec{\psi}_{-1}^{ref} + \overline{\mat{H}}_{L} \vec{\psi}_{0}^{ref} + \overline{\mat{H}}_{L,L} \vec{\psi}_{1}^{ref,(l)} = 0.
\end{equation}
Accordingly, we can see that the ratio of the SWFs at the left matching plane $\vec{\psi}_{1}^{ref,(l)} ( \vec{\psi}_{0}^{ref})^{-1}$ can be derived as
\begin{equation}
\label{eqn:04-12}
\vec{\psi}_{1}^{ref,(l)}\left( \vec{\psi}_{0}^{ref} \right)^{-1}= - (\overline{\mat{H}}_{L,L})^{-1}\left( \overline{\mat{H}}_{L} + \overline{\mat{H}}_{L,L}^\dag \vec{\psi}_{-1}^{ref} \left( \vec{\psi}_{0}^{ref} \right)^{-1} \right).
\end{equation}
In a similar way, the ratio at the right matching plane can be constructed as
\begin{equation}
\label{eqn:04-13}
\vec{\psi}_{n}^{(l)}\left( \vec{\psi}_{n+1} \right)^{-1} = - (\overline{\mat{H}}_{R,R}^\dag)^{-1}\left( \overline{\mat{H}}_{R} + \overline{\mat{H}}_{R,R} \vec{\psi}_{n+2}\left( \vec{\psi}_{n+1} \right)^{-1} \right).
\end{equation}
Note that, once the recursive equations Eqs.~(\ref{eqn:04-12}) and (\ref{eqn:04-13}) are solved, $\vec{\psi}_{1}^{ref,(l)} ( \vec{\psi}_{0}^{ref})^{-1}$ and $\vec{\psi}_{n}^{(l)}( \vec{\psi}_{n+1} )^{-1}$ contain part of the contribution of the rapidly varying evanescent wave excluded by the cutoff parameter $\lambda_{\text{min}}$. Inserting Eqs.~(\ref{eqn:04-12}) and (\ref{eqn:04-13}) into Eq.~(\ref{eqn:04-08}) and solving the simultaneous equation Eq.~(\ref{eqn:04-08}), we obtain the SWFs at the first layers of the left and right electrode regions $\vec{\psi}_{0}^{ref}$ and $\vec{\psi}_{n+1}$.

Defining the $m_{LL} \times m_{LL}$ ratio matrices\cite{PhysRevB_67_195315} as
\begin{eqnarray}
\label{eqn:04-14}
\mat{R}^{ref,(l+1)} &=& \vec{\psi}_{-1}^{ref}\left( \vec{\psi}_{0}^{ref} \right)^{-1}, \\
\label{eqn:04-15}
\mat{R}^{ref,(l)} &=& \vec{\psi}_{0}^{ref}\left( \vec{\psi}_{1}^{(l),ref} \right)^{-1}, \\
\label{eqn:04-16}
\mat{R}^{ref,(l-1)} &=& \vec{\psi}_{1}^{ref,(l)}\left( \vec{\psi}_{1}^{ref,(l-1)} \right)^{-1},
\end{eqnarray}
and the $m_{RR} \times m_{RR}$ ratio matrices as
\begin{eqnarray}
\label{eqn:04-17}
\mat{R}^{tra,(l+2)} &=& \vec{\psi}_{n+2}\left( \vec{\psi}_{n+1} \right)^{-1}, \\
\label{eqn:04-18}
\mat{R}^{tra,(l+1)} &=& \vec{\psi}_{n+1}\left( \vec{\psi}_{n}^{(l)} \right)^{-1}, \\
\label{eqn:04-19}
\mat{R}^{tra,(l)} &=& \vec{\psi}_{n}^{(l)}\left( \vec{\psi}_{n}^{(l-1)} \right)^{-1},
\end{eqnarray}
Eqs.~(\ref{eqn:04-12}) and (\ref{eqn:04-13}) can be rewritten in the form of continued fractions,\cite{PhysRevB_67_195315}
\begin{eqnarray}
\label{eqn:04-20}
\mat{R}^{ref,(l-1)} &=& - \left( \overline{\mat{H}}_{L} + \overline{\mat{H}}_{L,L}^\dag \mat{R}^{ref,(l)} \right)^{-1}\overline{\mat{H}}_{L,L}, \\
\label{eqn:04-21}
\mat{R}^{tra,(l)} &=& - \left( \overline{\mat{H}}_{R} + \overline{\mat{H}}_{R,R} \mat{R}^{tra,(l+1)} \right)^{-1}\overline{\mat{H}}_{R,R}^\dag.
\end{eqnarray}
Accordingly, the WFM formula Eq.~(\ref{eqn:04-08}) becomes
\begin{eqnarray}
\label{eqn:04-22}
\left(
\begin{array}{cc}
- \mat{G}^{(l)}_{1-l,1-l}\overline{\mat{H}}_{L,L}^\dag - \left( \mat{R}^{ref,(l)} \right)^{-1} & - \mat{G}^{(l)}_{1-l,n+l}\overline{\mat{H}}_{R,R} \\
- \mat{G}^{(l)}_{n+l,1-l}\overline{\mat{H}}_{L,L}^\dag & - \mat{G}^{(l)}_{n+l,n+l}\overline{\mat{H}}_{R,R} - \left( \mat{R}^{tra,(l+1)} \right)^{-1}
\end{array}
\right)
\left(
\begin{array}{c}
\vec{\psi}_{0}^{ref} \\
\vec{\psi}_{n+1}
\end{array}
\right) \nonumber \\
=
\left(
\begin{array}{c}
\mat{G}^{(l)}_{1-l,1-l}\overline{\mat{H}}_{L,L}^\dag \vec{\psi}_{0}^{in} + \vec{\psi}_{1}^{in,(l)}\\
\mat{G}^{(l)}_{n+l,1-l}\overline{\mat{H}}_{L,L}^\dag \vec{\psi}_{0}^{in}
\end{array}
\right).
\end{eqnarray}
The SWFs in the transition region are obtained by substituting the obtained $\vec{\psi}_{0}^{ref}$ and $\vec{\psi}_{n+1}$ into Eq.~(\ref{eqn:04-02}). Note that only parts of the block-matrix elements of the Green's functions matrix $\mat{G}^{(l)}_{i,1-l}$ and $\mat{G}^{(l)}_{i,n+l}$ $(i=1-l, \cdots,n+l)$ are needed to obtain the SWFs. 

Furthermore, using the ratio matrices, the self-energy terms of the electrodes are expressed as
\begin{eqnarray}
\label{eqn:04-23}
\mat{\Sigma}_L^{(l)} &=& -\overline{\mat{H}}_{L,L}^\dag \mat{R}^{ref,(l)}, \\
\label{eqn:04-24}
\mat{\Sigma}_R^{(l)} &=& -\overline{\mat{H}}_{R,R} \mat{R}^{tra,(l+1)},
\end{eqnarray}
and the SWFs in the electrode regions are rewritten as
\begin{eqnarray}
\label{eqn:04-26}
\vec{\psi}_{0} &=& \vec{\psi}_{0}^{ref}+\vec{\psi}_{0}^{in} \nonumber \\
&=& \vec{\psi}_{0}^{ref} \left( \vec{\psi}_{1}^{ref,(l)} \right)^{-1} \vec{\psi}_{1}^{ref,(l)} + \vec{\psi}_{0}^{in} \nonumber \\
&=& \mat{R}^{ref,(l)} \left( \vec{\psi}_{1}^{(l)} - \vec{\psi}_{1}^{in,(l)} \right)+\vec{\psi}_{0}^{in}, \\
\label{eqn:04-27}
\vec{\psi}_{n+1} &=& \vec{\psi}_{n+1} \left( \vec{\psi}_{n}^{(l)} \right)^{-1} \vec{\psi}_{n}^{(l)} \nonumber \\
&=& \mat{R}^{tra,(l)}\vec{\psi}_{n}^{(l)}.
\end{eqnarray}
Inserting Eqs.~(\ref{eqn:04-23}), (\ref{eqn:04-24}), (\ref{eqn:04-26}), and (\ref{eqn:04-27}) into Eq.~(\ref{eqn:04-02}) yields\cite{PhysRevB_76_235422}
\begin{eqnarray}
\label{eqn:04-28}
E\widehat{\mat{S}}^{(l)}-\widehat{\mat{H}}^{(l)}-\widetilde{\mat{H}}^{(l)}
\left(
\begin{array}{c}
\vec{\psi}_{1}^{(l)} \\
\vdots \\
\vdots \\
\vdots \\
\vec{\psi}_{n}^{(l)}
\end{array}
\right)
=
\left(
\begin{array}{c}
-\overline{\mat{H}}_{L,L}^\dag\vec{\psi}_{0}^{in} - \mat{\Sigma}_L^{(l)}\vec{\psi}_{1}^{in,(l)} \\
0 \\
\vdots \\
0 \\
0
\end{array}
\right), 
\end{eqnarray}
where $\widetilde{\mat{H}}^{(l)}$ is a zero matrix except that the upper left $m_{LL} \times m_{LL}$ and lower right $m_{RR} \times m_{RR}$ block-matrix elements are $\mat{\Sigma}_L^{(l)}$ and $\mat{\Sigma}_R^{(l)}$, respectively:
\begin{eqnarray}
\label{eqn:04-29}
\widetilde{\mat{H}}^{(l)}= \left(
\begin{array}{ccccc}
\mat{\Sigma}_L^{(l)} & 0 & \cdots & \cdots & 0 \\
0 & 0 & & & \vdots \\
\vdots & & \ddots & & \vdots \\
\vdots & & & 0 & 0 \\
0 & \cdots & \cdots & 0 & \mat{\Sigma}_R^{(l)} \\
\end{array}
\right).
\end{eqnarray}
The SWFs can be evaluated without computing the Green's functions of the isolated transition region $\widehat{\mat{G}}^{(l)}$ if one solves Eq.~(\ref{eqn:04-28}). Inserting $\vec{\psi}_{1}^{(l)}$ and $\vec{\psi}_{n}^{(l)}$ into Eqs.~(\ref{eqn:04-26}) and (\ref{eqn:04-27}), one has the SWFs at the first layers of the electrode regions $\vec{\psi}^{ref}_{0}(=\vec{\psi}_{0}-\vec{\psi}^{in}_{0})$ and $\vec{\psi}_{n+1}$, which are used to compute the transmission and reflection coefficients.

\subsection{Moving matching plane of wave function matching formula}
\label{sec:Moving matching plane of wave function matching formula}
The extension of the transition region requires additional computations because the number of dimensions of the Green's functions $\widehat{\mat{G}}^{(l)}$ and the number of variables in the simultaneous equations Eq.~(\ref{eqn:04-28}) linearly increase with respect to the number of inserted extra layers. In this subsection, we describe the procedure to move the matching planes to the inside of the transition region without loss of accuracy. In the previous subsection, the left (right) matching plane is set between $\vec{\psi}_{0}$ and $\vec{\psi}_{1}^{(l)}$ ($\vec{\psi}_{n}^{(l)}$ and $\vec{\psi}_{n+1}$). We are going to set the left (right) matching plane between $\vec{\psi}_{1}^{(l)}$ and $\vec{\psi}_{1}^{(l-1)}$ ($\vec{\psi}_{n}^{(l-1)}$ and $\vec{\psi}_{n}^{(l-1)}$). Khomyakov {\it et al.} proved the translational invariance of the total transmission probability with respect to moving the matching planes to {\it the inside of the electrode regions}.\cite{PhysRevB_70_195402} However, when the matching planes are moved to {\it the inside of the transition region}, we need to treat the rapidly decreasing evanescent waves, which are excluded from the electrode regions by introducing the cutoff parameter $\lambda_{\text{min}}$ but contribute to the SWFs in the transition region.

Using the Kohn-Sham Hamiltonian for the model with $l-1$ extra layers, the Hamiltonian for the model with $l$ extra layers can be rewritten as
\begin{equation}
\label{eqn:05-01}
E\widehat{\mat{S}}^{(l)}-\widehat{\mat{H}}^{(l)}=\left(
\begin{array}{ccc}
E\mat{S}_L-\mat{H}_L & \mat{A}^{(l)}_L & \vec{0} \\
\mat{A}^{(l) \dag}_L & E\widehat{\mat{S}}^{(l-1)}-\widehat{\mat{H}}^{(l-1)} & \mat{A}^{(l)}_R \\
\vec{0} & \mat{A}^{(l) \dag}_R & E\mat{S}_R-\mat{H}_R
\end{array}
\right)=\left(
\begin{array}{ccc}
\overline{\mat{H}}_{L} & \mat{A}^{(l)}_L & \vec{0} \\
\mat{A}^{(l) \dag}_L & E\widehat{\mat{S}}^{(l-1)}-\widehat{\mat{H}}^{(l-1)} & \mat{A}^{(l)}_R \\
\vec{0} & \mat{A}^{(l) \dag}_R & \overline{\mat{H}}_{R}
\end{array}
\right),
\end{equation}
where
\begin{eqnarray}
\label{eqn:05-02}
\mat{A}^{(l)}_L &=& (\overline{\mat{H}}_{L,L},\vec{0},\cdots,\vec{0}), \\
\label{eqn:05-03}
\mat{A}^{(l)}_R &=& (\overline{\mat{H}}_{R,R},\vec{0},\cdots,\vec{0}).
\end{eqnarray}
$\mat{A}^{(l)}_L(\mat{A}^{(l)}_R)$ is the $m_L \times (m_T+(l-1)(m_L+m_R))[m_R \times (m_T+(l-1)(m_L+m_R))]$ zero matrix except that the $m_L \times m_L(m_R \times m_R)$ block-matrix element at the left edge is $\overline{\mat{H}}_{L,L}(\overline{\mat{H}}_{R,R})$. Consequently, Eq.~(\ref{eqn:04-01}) becomes
\begin{equation}
\label{eqn:05-04}
\left(
\begin{array}{c|c|c}
\overline{\mat{H}}_{L} & \mat{A}^{(l)}_L & \vec{0} \\
\mat{A}^{(l) \dag}_L & E\widehat{\mat{S}}^{(l-1)}-\widehat{\mat{H}}^{(l-1)} & \mat{A}^{(l)}_R \\
\vec{0} & \mat{A}^{(l) \dag}_R & \overline{\mat{H}}_{R}
\end{array}
\right)
\left(
\begin{array}{c}
\vec{\psi}_{1}^{(l)} \\
\hline
\vec{\psi}_{1}^{(l-1)} \\
\vdots \\
\vec{\psi}_{n}^{(l-1)} \\
\hline
\vec{\psi}_{n}^{(l)}
\end{array}
\right)
=
\left(
\begin{array}{c}
-\overline{\mat{H}}_{L,L}^\dag\vec{\psi}_{0} \\
\hline
0 \\
\vdots \\
0 \\
\hline
-\overline{\mat{H}}_{R,R}\vec{\psi}_{n+1}
\end{array}
\right).
\end{equation}
Eliminating the first and last block rows yields
\begin{equation}
\label{eqn:05-05}
E\widehat{\mat{S}}^{(l-1)}-\widehat{\mat{H}}^{(l-1)}
\left(
\begin{array}{c}
\vec{\psi}_{1}^{(l-1)} \\
\vdots \\
\vdots \\
\vdots \\
\vec{\psi}_{n}^{(l-1)}
\end{array}
\right)
=
\left(
\begin{array}{c}
-\overline{\mat{H}}_{L,L}^\dag\vec{\psi}_{1}^{(l)} \\
0 \\
\vdots \\
0 \\
-\overline{\mat{H}}_{R,R}\vec{\psi}_{n}^{(l)}
\end{array}
\right).
\end{equation}
Using $\widehat{\mat{G}}^{(l-1)}$, Eq.~(\ref{eqn:05-05}) can be rewritten as
\begin{equation}
\label{eqn:05-06}
\left(
\begin{array}{c}
\vec{\psi}_{1}^{(l-1)} \\
\vdots \\
\vdots \\
\vdots \\
\vec{\psi}_{n}^{(l-1)}
\end{array}
\right)
=
\widehat{\mat{G}}^{(l-1)}\left(
\begin{array}{c}
-\overline{\mat{H}}_{L,L}^\dag\vec{\psi}_{1}^{(l)} \\
0 \\
\vdots \\
0 \\
-\overline{\mat{H}}_{R,R}\vec{\psi}_{n}^{(l)}
\end{array}
\right).
\end{equation}
In analogy to Eq.~(\ref{eqn:04-01}), we then have a WFM formula where the matching planes are shifted by one layer inside of the transition region.
\begin{eqnarray}
\label{eqn:05-07}
\left(
\begin{array}{cc}
- \mat{G}^{(l-1)}_{2-l,2-l}\overline{\mat{H}}_{L,L}^\dag - \left( \mat{R}^{ref,(l-1)} \right)^{-1} & - \mat{G}^{(l-1)}_{2-l,n+l-1}\overline{\mat{H}}_{R,R} \\
- \mat{G}^{(l-1)}_{n+l-1,2-l}\overline{\mat{H}}_{L,L}^\dag & - \mat{G}^{(l-1)}_{n+l-1,n+l-1}\overline{\mat{H}}_{R,R} - \left( \mat{R}^{tra,(l)} \right)^{-1}
\end{array}
\right)
\left(
\begin{array}{c}
\vec{\psi}_{1}^{ref,(l)} \\
\vec{\psi}_{n}^{(l)}
\end{array}
\right) \nonumber \\
=
\left(
\begin{array}{c}
\mat{G}^{(l-1)}_{2-l,2-l}\overline{\mat{H}}_{L,L}^\dag \vec{\psi}_{1}^{in,(l)} + \vec{\psi}_{1}^{in,(l-1)}\\
\mat{G}^{(l-1)}_{n+l-1,2-l}\overline{\mat{H}}_{L,L}^\dag \vec{\psi}_{1}^{in,(l)}
\end{array}
\right).
\end{eqnarray}
Substituting $\mat{R}^{ref,(l)}$ and $\mat{R}^{tra,(l+1)}$ into Eqs.~(\ref{eqn:04-20}) and (\ref{eqn:04-21}), respectively, we obtain $\mat{R}^{ref,(l-1)}$ and $\mat{R}^{tra,(l)}$. Although the Green's function $\widehat{\mat{G}}^{(l-1)}$ is required, its dimension is reduced from $m_T+l(m_L+m_R)$ to $ m_T+(l-1)(m_L+m_R)$.

By repeating the shift of the matching planes by using Eqs.~(\ref{eqn:04-22}) and (\ref{eqn:05-07}) and solving Eqs.~(\ref{eqn:04-20}) and (\ref{eqn:04-21}) $l$ times, the matching plane gets set between $\vec{\psi}_{1}^{ref,(1)}$ and $\vec{\psi}_{1}^{ref,(0)}$ ($\vec{\psi}_{n}^{(0)}$ and $\vec{\psi}_{n}^{(1)}$) for the left (right) electrode region. The WFM formula is expressed as
\begin{eqnarray}
\label{eqn:05-08}
\left(
\begin{array}{cc}
- \mat{G}^{(0)}_{1,1}\overline{\mat{H}}_{L,L}^\dag - \left( \mat{R}^{ref,(0)} \right)^{-1} & - \mat{G}^{(0)}_{1,n}\overline{\mat{H}}_{R,R} \\
- \mat{G}^{(0)}_{n,1}\overline{\mat{H}}_{L,L}^\dag & - \mat{G}^{(0)}_{n,n}\overline{\mat{H}}_{R,R} - \left( \mat{R}^{tra,(1)} \right)^{-1}
\end{array}
\right)
\left(
\begin{array}{c}
\vec{\psi}_{1}^{ref,(1)} \\
\vec{\psi}_{n}^{(1)}
\end{array}
\right) \nonumber \\
=
\left(
\begin{array}{c}
\mat{G}^{(0)}_{1,1}\overline{\mat{H}}_{L,L}^\dag \vec{\psi}_{1}^{in,(1)} + \vec{\psi}_{1}^{in,(0)}\\
\mat{G}^{(0)}_{n,1}\overline{\mat{H}}_{L,L}^\dag \vec{\psi}_{1}^{in,(1)}
\end{array}
\right).
\end{eqnarray}
The SWFs in the transition region are computed by substituting $\vec{\psi}_{1}^{ref,(1)}$ and $\vec{\psi}_{n}^{(1)}$ into Eq.~(\ref{eqn:05-06}).

If we do not use the Green's function of the transition region $\widehat{\mat{G}}^{(0)}$, we solve the following simultaneous equations corresponding to Eq.~(\ref{eqn:04-28}):
\begin{eqnarray}
\label{eqn:05-10}
E\widehat{\mat{S}}^{(0)}-\widehat{\mat{H}}^{(0)}-\widetilde{\mat{H}}^{(0)}
\left(
\begin{array}{c}
\vec{\psi}_{1}^{(0)} \\
\vdots \\
\vdots \\
\vdots \\
\vec{\psi}_{n}^{(0)}
\end{array}
\right)
=
\left(
\begin{array}{c}
-\overline{\mat{H}}_{L,L}^\dag\vec{\psi}_{1}^{in,(1)} - \mat{\Sigma}_L^{(0)} \vec{\psi}_{1}^{in,(0)} \\
0 \\
\vdots \\
0 \\
0
\end{array}
\right),
\end{eqnarray}
where
\begin{eqnarray}
\label{eqn:05-11}
\widetilde{\mat{H}}^{(0)}= \left(
\begin{array}{ccccc}
\mat{\Sigma}_L^{(0)} & 0 & \cdots & \cdots & 0 \\
0 & 0 & & & \vdots \\
\vdots & & \ddots & & \vdots \\
\vdots & & & 0 & 0 \\
0 & \cdots & \cdots & 0 & \mat{\Sigma}_R^{(0)} \\
\end{array}
\right).
\end{eqnarray}
The SWFs of the electrode side of the matching planes are obtained from Eqs.~(\ref{eqn:04-26}) and (\ref{eqn:04-27}).

This result implies that the translational invariance of the SWFs with respect to moving the matching planes to the inside of the transition region is preserved even when the rapidly varying evanescent waves are excluded. In addition, when the sum of the number of iterations for Eqs.~(\ref{eqn:04-20}) and (\ref{eqn:04-21}) and the number of the extra layers are identical, exactly the same SWFs can be obtained as long as the boundary condition is properly imposed, as mentioned in Sec.~\ref{sec:Boundary condition}. In Sec.~\ref{sec:Translational invariance of the transmission probability with respect to moving matching plane}, we demonstrate that the translational invariance is not retained when the transition region and electrode regions overlap as shown in Table~\ref{tbl:1} and the set of generalized Bloch waves is incomplete. 

Finally, let us consider the convergence of Eqs.~(\ref{eqn:04-20}) and (\ref{eqn:04-21}). Suppose that the solutions of Eq.~(\ref{eqn:05-08}) with $l$ extra layers are $\vec{\psi}_{1}^{ref,(1)}$ and $\vec{\psi}_{n}^{(1)}$ and the solutions with $l+1$ extra layers are $\vec{\psi}_{1}^{'ref,(1)}$ and $\vec{\psi}_{n}^{'(1)}$, $\vec{\psi}_{1}^{ref,(1)}= \vec{\psi}_{1}^{'ref,(1)}$ and $\vec{\psi}_{n}^{(1)}= \vec{\psi}_{n}^{'(1)}$ when $l$ is large enough. The first row of Eq.~(\ref{eqn:05-08}) for $l$ extra layers is
\begin{eqnarray}
\label{eqn:05-12}
- \mat{G}^{(0)}_{1,1}\overline{\mat{H}}_{L,L}^\dag\vec{\psi}_{1}^{ref,(1)} - \left( \mat{R}^{ref,(0)} \right)^{-1}\vec{\psi}_{1}^{ref,(1)} - \mat{G}^{(0)}_{1,n}\overline{\mat{H}}_{R,R} \vec{\psi}_{n}^{(1)} &=& \mat{G}^{(0)}_{1,1}\overline{\mat{H}}_{L,L}^\dag \vec{\psi}_{1}^{in,(1)}  + \vec{\psi}_{1}^{in,(0)}. 
\end{eqnarray}
Inserting $\vec{\psi}_{1}^{ref,(1)}$ and $\vec{\psi}_{n}^{(1)}$ into Eq.~(\ref{eqn:05-08}) for $l+1$ extra layers, we have
\begin{eqnarray}
\label{eqn:05-13}
- \mat{G}^{(0)}_{1,1}\overline{\mat{H}}_{L,L}^\dag\vec{\psi}_{1}^{ref,(1)} - \left( \mat{R}^{'ref,(0)} \right)^{-1}\vec{\psi}_{1}^{ref,(1)} - \mat{G}^{(0)}_{1,n}\overline{\mat{H}}_{R,R} \vec{\psi}_{n}^{(1)} &=& \mat{G}^{(0)}_{1,1}\overline{\mat{H}}_{L,L}^\dag \vec{\psi}_{1}^{in,(1)}  + \vec{\psi}_{1}^{in,(0)},
\end{eqnarray}
where $\mat{R}^{'ref,(0)}$ is the ratio matrix of the computational model with $l+1$ extra layers. Subtracting Eq.~(\ref{eqn:05-13}) from Eq.~(\ref{eqn:05-12}), we obtain
\begin{equation}
\label{eqn:05-14}
\left( \mat{R}^{ref,(0)} \right)^{-1}=\left( \mat{R}^{'ref,(0)} \right)^{-1}.
\end{equation}
$\mat{R}^{ref,(0)}$ is computed by solving Eq.~(\ref{eqn:04-20}) $l+1$ times with the initial matrix of $(\tilde{\mat{\Lambda}}_{L}^{-})^{-1}\tilde{\mat{\Phi}}_{L}^{-}( \tilde{\mat{\Phi}}_{L}^{-} )^{-1}$, while $\mat{R}^{'ref,(0)}$ is obtained by solving Eq.~(\ref{eqn:04-20}) $l+2$ times with the same initial matrix. This indicates that $\mat{R}^{ref,(0)}$ is uniquely determined when $l$ is sufficiently large. In other words, by repeatedly solving Eqs.~(\ref{eqn:04-20}) and (\ref{eqn:04-21}) until $\mat{R}^{ref,(0)}$ and $\mat{R}^{tra,(1)}$ become consistent, we obtain $\mat{R}^{ref,(0)}$ and $\mat{R}^{tra,(1)}$ for $l \rightarrow \infty$ and the SWFs for a truly semi-infinite system. The convergence behavior of the ratio matrix is demonstrated in Appendix~\ref{sec:Convergence behavior of the ratio matrix}.

\subsection{Transmission probability}
\label{sec:Transmission probability}
The expression of the conductance at zero bias limit is given by the Landauer-B\"uttiker formula.\cite{PhysRevB_31_006207} The Landauer-B\"uttiker formula using the transmission coefficient matrix is
\begin{equation}
\label{eqn:07-01}
G(E)=\frac{2e^2}{h}\mbox{Tr}\mat{T}=\frac{2e^2}{h}\mbox{Tr}\left[\left( \mat{v}^{in} \right)^{-1}\mat{t}^\dag \mat{v}^{tra} \mat{t}\right],
\end{equation}
where $\mat{t}$ is an $m_{RR} \times m_k$ transmission coefficient matrix with $m_k$ being the number of the incident waves, and $\mat{v}^{in}$($\mat{v}^{tra}$) is an $m_k$-($m_{RR}$-)dimensional matrix describing the group velocity of the incident (transmission) waves. In Ref.~\onlinecite{PhysRevB_86_195406}, the group velocity of the incident waves is written as
\begin{eqnarray}
\label{eqn:07-02}
\mat{v}^{in}&=&{\mathrm i}a\mat{\Phi}_L^{+ \dag} \mat{\Gamma}_L^{(0)} \mat{\Phi}_L^+,
\end{eqnarray}
where
\begin{eqnarray}
\label{eqn:07-03}
\mat{\Phi}_L^+ &=& \left\{ \phi_{L,1}^+, \cdots, \phi_{L,m_k}^+ \right\}.
\end{eqnarray}
The matrix for the group velocity is diagonal when the self-energy terms of electrodes are exactly obtained. However, due to the exclusion of the rapidly varying evanescent waves and numerical error in the computation of the Green's functions, the off-diagonal elements do not vanish completely. The group velocity of the transmitted waves is expressed as
\begin{eqnarray}
\label{eqn:07-04}
\mat{v}^{tra}&=&{\mathrm i}a(\widetilde{\mat{\Phi}}_{R}^+)^\dag \mat{\Gamma}_{R}^{(0)} \widetilde{\mat{\Phi}}_{R}^+.
\end{eqnarray}
Here, $a$ is the length of the unit cell and
\begin{equation}
\label{eqn:07-05}
\mat{\Gamma}_L^{(l)}={\mathrm i} \left( \mat{\Sigma}_L^{(l)} - \mat{\Sigma}_L^{(l) \dag} \right).
\end{equation}
From Eq.~(\ref{eqn:03-06}), the transmission coefficient matrix is obtained by
\begin{eqnarray}
\label{eqn:07-06}
\vec{t}_k &=& \left(
\begin{array}{c}
\tilde{\mat{\Phi}}_{R}^{+}, \mathring{\mat{\Phi}}_{R}^{+}
\end{array}
\right)^{-1}\vec{\psi}_{n,k}^{(1)}
\approx
\left(
\begin{array}{c}
\tilde{\mat{\Phi}}_{R}^{+}, \mathring{\mat{0}}
\end{array}
\right)^{-1}\vec{\psi}_{n,k}^{(1)},
\end{eqnarray}
and
\begin{eqnarray}
\label{eqn:07-07}
\mat{t}&=& (\vec{t}_1, \cdots, \vec{t}_{m_k} ).
\end{eqnarray}
Khomyakov {\it et al.} and S\o{}rensen {\it et al.} instructed one to use the pseudoinverses for ($\tilde{\mat{\Phi}}_{R}^{+}, \mathring{\mat{0}}$).\cite{PhysRevB_70_195402,PhysRevB_79_205322} However, the SWFs $\vec{\psi}_{n,k}^{(1)}$ include the components of $\mathring{\mat{\Phi}}_{R}^{+}$, because they are determined on real-space grids or linear combinations of bases in the transition region. In addition, as we show in Appendix~\ref{sec:Deterioration of the translational invariance of the transmission probability due to the pseudoinverse}, translational invariance is not retained when the pseudoinverses are used in the computation of the transmission coefficients in Eq.~(\ref{eqn:07-06}). Therefore, we introduce $\mat{\Phi}_{R}^{' +}$ so that
\begin{equation}
\label{eqn:07-08}
\check{\mat{\Phi}}_{R}^{+}=\left(
\begin{array}{c}
\tilde{\mat{\Phi}}_{R}^{+}, \mat{\Phi}_{R}^{' +}
\end{array}
\right)
\end{equation}
becomes a regular matrix. Using $\check{\mat{\Phi}}_{R}^{+}$, we define
\begin{eqnarray}
\label{eqn:07-09}
\check{\mat{v}}^{tra}&=&{\mathrm i}a(\check{\mat{\Phi}}_{R}^+)^\dag \mat{\Gamma}_{R}^{(0)} \check{\mat{\Phi}}_{R}^+, \\
\label{eqn:07-10}
\check{\vec{t}}_k &=& \left( \check{\mat{\Phi}}_{R}^{+} \right)^{-1}\vec{\psi}_{n,k}^{(1)}, \\
\label{eqn:07-11}
\check{\mat{t}}&=& (\check{\vec{t}}_1, \cdots, \check{\vec{t}}_{m_k} ).
\end{eqnarray}
Substituting Eqs.~(\ref{eqn:07-09}), (\ref{eqn:07-10}), and (\ref{eqn:07-11}), we have
\begin{equation}
\label{eqn:07-12}
\mat{T}=\left( \mat{v}^{in} \right)^{-1}\check{\mat{t}}^\dag \check{\mat{v}}^{tra} \check{\mat{t}}=\left( \mat{v}^{in} \right)^{-1}\mat{t}^\dag \mat{v}^{tra} \mat{t}.
\end{equation}
The transmission matrix is invariant when one takes $\mat{\Phi}_{R}^{' +}$ so that $\check{\mat{\Phi}}_{R}^{+}$ is a regular matrix. The proof is given in Appendix~\ref{sec:Invariance of the transmission matrix with respect to the transmission waves}. Very often one is only interested in the conductance. In that case one can skip the computation of the transmission coefficients. The $k$th diagonal element of the transmission matrix $T_{k,k}$ is given by
\begin{eqnarray}
\label{eqn:07-13}
T_{k,k}&=&\left( v_{k,k}^{in} \right)^{-1}\check{\vec{t}}_k^\dag \check{\mat{v}}^{tra} \check{\vec{t}}_k \nonumber \\
&=&{\mathrm i}a\left( v^{in}_{k,k} \right)^{-1} \left[ \left(\check{\mat{\Phi}}_{R}^{+} \right)^{-1} \vec{\psi}_{n,k}^{(1)} \right]^\dag \left(\check{\mat{\Phi}}_{R}^{+} \right)^\dag \mat{\Gamma}_{R}^{(0)} \check{\mat{\Phi}}_{R}^{+} \left(\check{\mat{\Phi}}_{R}^{+} \right)^{-1} \vec{\psi}_{n,k}^{(1)} \nonumber \\
&=&{\mathrm i}a\left( v^{in}_{k,k} \right)^{-1} \vec{\psi}_{n,k}^{(1) \dag} \mat{\Gamma}_{R}^{(0)} \vec{\psi}_{n,k}^{(1)},
\end{eqnarray}
where $v_{i,j}^{in}$ is the $( i, j)$th element of $\mat{v}^{in}$. Analogous to the transmission matrix and probability, the reflection matrix and probability can be obtained from $\vec{\psi}_{1,k}^{ref, (1)}$. 
Note that the transmission probability is given only from $\vec{\psi}_{n,k}^{(1) \dag}$ and $\mat{\Gamma}_{R}^{(0)}$. 
This is contrastive to the nonequilibrium Green's function method, which uses the Green's function of the transition region being expensive to be calculated. 
Section~\ref{sec:Accuracy of the transmission probability obtained by SWF} describes the efficiency of using the inverse of regular a matrix $\check{\mat{\Phi}}_{R}^{+}$ instead of the pseudoinverse of $\widetilde{\mat{\Phi}}_{R}^{+}$.

\section{Numerical example}
\subsection{Translational invariance of transmission probability with respect to moving matching plane}
\label{sec:Translational invariance of the transmission probability with respect to moving matching plane}
S\o{}rensen {\it et al.} proposed a method that extends the transition region by inserting extra layers to improve the accuracy of the WFM technique.\cite{PhysRevB_79_205322} Insertion of the extra layers corresponds to moving the matching plane to the inside of the electrode region. From the problem of the penetration of quantum particles into a one-dimensional square potential barrier on the basis of quantum mechanics, it is obvious that the transmission probability does not depend on the position of the matching plane. Translational invariance is retained as long as all the generalized Bloch waves are included for the WFM procedure. When the rapidly decreasing evanescent waves are excluded by the cutoff parameter $\lambda_{\text{min}}$, it is not trivial whether the invariance is kept at the boundary between the electrode and transition regions. Khomyakov {\it et al.} proved the translational invariance with respect to moving the matching plane toward the electrode region.\cite{PhysRevB_70_195402} Moving the matching plane to the electrode side is straightforward because the excluded evanescent waves do not contribute to the SWFs anymore. When the planes are moved into the transition region, one needs to consider the contribution of the excluded evanescent waves because the waves do not vanish in the transition region.

\begin{figure}[htb]
\begin{center}
\includegraphics{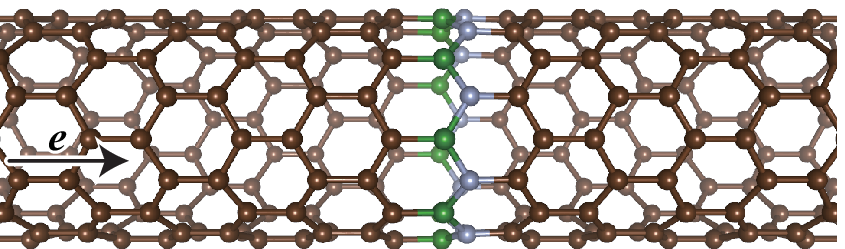}
\caption{(Color online) Computational model where one BN zigzag ring is sandwiched between (9,0) CNT electrodes. Large dark, small dark, and small light balls are N, C, and B atoms, respectively. \label{fig:2}}
\end{center}
\end{figure}

As an example of a practical calculation, we will examine the transport properties of a (9,0) B--N nanotube sandwiched between (9,0) carbon nanotubes (CNTs). Figure~\ref{fig:2} shows the computational model. The $x$- and $y$-axes are the directions perpendicular to the nanotube, and the $z$-axis is parallel to the nanotube. A valence electron-ion interaction is treated by norm-conserving pseudopotentials generated by the scheme proposed by Troullier and Martins.\cite{PhysRevB_43_001993,norm} The local density approximation\cite{PhysRevB_23_005048} of the density functional theory\cite{PhysRev_136_B864,RevModPhys_71_001253} is used to describe the exchange and correlation effects. We use the real-space finite-difference method for the first-principles calculation implemented in the first-principles calculation code {\sc rspace},\cite{PhysRevLett_82_005016,PhysRevB_72_085115,PhysRevB_82_205115,icp} which enables us to calculate the transport properties of nanostructures between the semi-infinite electrodes. In the real-space grid formalism, $m_L$ ($m_R$) is not equal to $m_{LL}$ ($m_{RR}$). For that reason, the procedure described in Appendix~\ref{sec:Wave function matching for general case} is used. The central finite-difference formula\cite{PhysRevLett_72_001240} ($N=1$ in Ref.~\onlinecite{PhysRevB_50_011355}) is used for the second-order derivation arising from the kinetic-energy operator in the Kohn-Sham equation. A conventional supercell under periodic boundary conditions in all directions with a real-space grid spacing of $\sim$ 0.24 \AA \hspace{2mm} is used to determine the Kohn-Sham effective potential. The dimensions of the supercell are set as $L_x=13.34$ \AA, $L_y=13.34$ \AA, and $L_z=4.32$ \AA \hspace{2mm} for the electrode regions and $L_x=13.34$ \AA, $L_y=13.34$ \AA, and $L_z=8.64$ \AA \hspace{2mm} for the transition region. Here, $L_x$ and $L_y$ are the lateral lengths of the supercell in the $x$- and $y$-directions, respectively, and $L_z$ is the length in the $z$-direction. Table~\ref{tbl:3} shows the transmission probability with respect to the number of extra layers $l$ and the number of iterations for the continued-fraction equations (Eqs.~(\ref{eqn:06-19}) and (\ref{eqn:06-20})) $N_{it}$. Since all the WFM methods proposed so far use the same boundary conditions between the electrode and transition regions, we use the method proposed by Fujimoto and Hirose.\cite{PhysRevB_67_195315} To eliminate the numerical error caused by the pseudoinverses, all the generalized Bloch waves are included in $\mat{\Phi}_{R}^{+}$ for Eq.~(\ref{eqn:07-06}). In the conventional WFM formalisms, the computation of the continued-fraction equations is skipped when the extra layer is not inserted because $\vec{\psi}_{1}^{ref,(l)}(\xi_1)( \vec{\psi}_{0}^{ref}(\xi_3) )^{-1}$ and $\vec{\psi}_{n}^{(l)}(\xi_3)( \vec{\psi}_{n+1}(\xi_1) )^{-1}$ in Eq.~(\ref{eqn:06-05}) can be obtained as $\tilde{\mat{\Phi}}_{L}^{-}(\xi_1) ( (\tilde{\mat{\Lambda}}_{L}^{-})^{-1}\tilde{\mat{\Phi}}_{L}^{-}(\xi_3) )^{-1}$ and $\tilde{\mat{\Phi}}_{R}^{+}(\xi_3) ( \tilde{\mat{\Lambda}}_{R}^{+}\tilde{\mat{\Phi}}_{R}^{+}(\xi_1) )^{-1}$, respectively, owing to the overlap of the layers in the transition and electrode regions. One can see that while the translational invariance deteriorates between $N_{it}=0$ and $N_{it}=1$ in the conventional WFM methods, it is retained nicely in the present method. The translational invariance is destroyed by the inconsistency of the SWFs at the overlapping layers; the deterioration is suppressed when $\lambda_{\text{min}}$ is small or $l$ is large.

From these results, we can conclude that the overlap between the electrode and transition region should be removed to maintain the translational invariance of the transmission probability. However, the numerical error caused by the deterioration of the translational invariance of the transmission probability is small and is not the origin of the degradation of the accuracy when the transition layer is not extended as reported in Ref.~\onlinecite{PhysRevB_79_205322}.

\begin{table}
\caption{Transmission probability with respect to the number of extra layers $l$ and number of iterations for the continued-fraction equations $N_{it}$ for (9,0) B--N nanotube between (9,0) CNTs. \label{tbl:3}}
\centering
\begin{tabular}{ccccc}
\hline \hline 
$\lambda_{\text{min}}$&$l$ & $N_{it}$ & conventional method & present method \\ \hline
0.90 & 3 & 0  & 0.52046191 & 0.52046575 \\
     & 2 & 1  & 0.52045862 & 0.52046575 \\
     & 1 & 2  & 0.52045862 & 0.52046575 \\
     & 0 & 3  & 0.52045862 & 0.52046575 \\
&&&& \\
     & 2 & 0  & 0.52044949 & 0.52045862 \\
     & 1 & 1  & 0.52044135 & 0.52045862 \\
     & 0 & 2  & 0.52044135 & 0.52045862 \\
&&&& \\
     & 1 & 0  & 0.52041587 & 0.52044135 \\
     & 0 & 1  & 0.52039468 & 0.52044135 \\
&&&& \\
0.65 & 3 & 0  & 0.52046511 & 0.52046809 \\
     & 2 & 1  & 0.52046378 & 0.52046809 \\
     & 1 & 2  & 0.52046378 & 0.52046809 \\
     & 0 & 3  & 0.52046378 & 0.52046809 \\
&&&& \\
     & 2 & 0  & 0.52045659 & 0.52046378 \\
     & 1 & 1  & 0.52045277 & 0.52046378 \\
     & 0 & 2  & 0.52045277 & 0.52046378 \\
&&&& \\
     & 1 & 0  & 0.52043133 & 0.52045277 \\
     & 0 & 1  & 0.52042062 & 0.52045277 \\
&&&& \\
0.10 & 2 & 0  & 0.52047069 & 0.52047070 \\
     & 1 & 1  & 0.52047070 & 0.52047070 \\
     & 0 & 2  & 0.52047070 & 0.52047070 \\
&&&& \\
     & 1 & 0  & 0.52046908 & 0.52047070 \\
     & 0 & 1  & 0.52047028 & 0.52047070 \\
\hline \hline 
\end{tabular}
\end{table}

\subsection{Accuracy of the transmission matrix and probability obtained by SWF}
\label{sec:Accuracy of the transmission probability obtained by SWF}
\begin{figure}[htb]
\begin{center}
\includegraphics{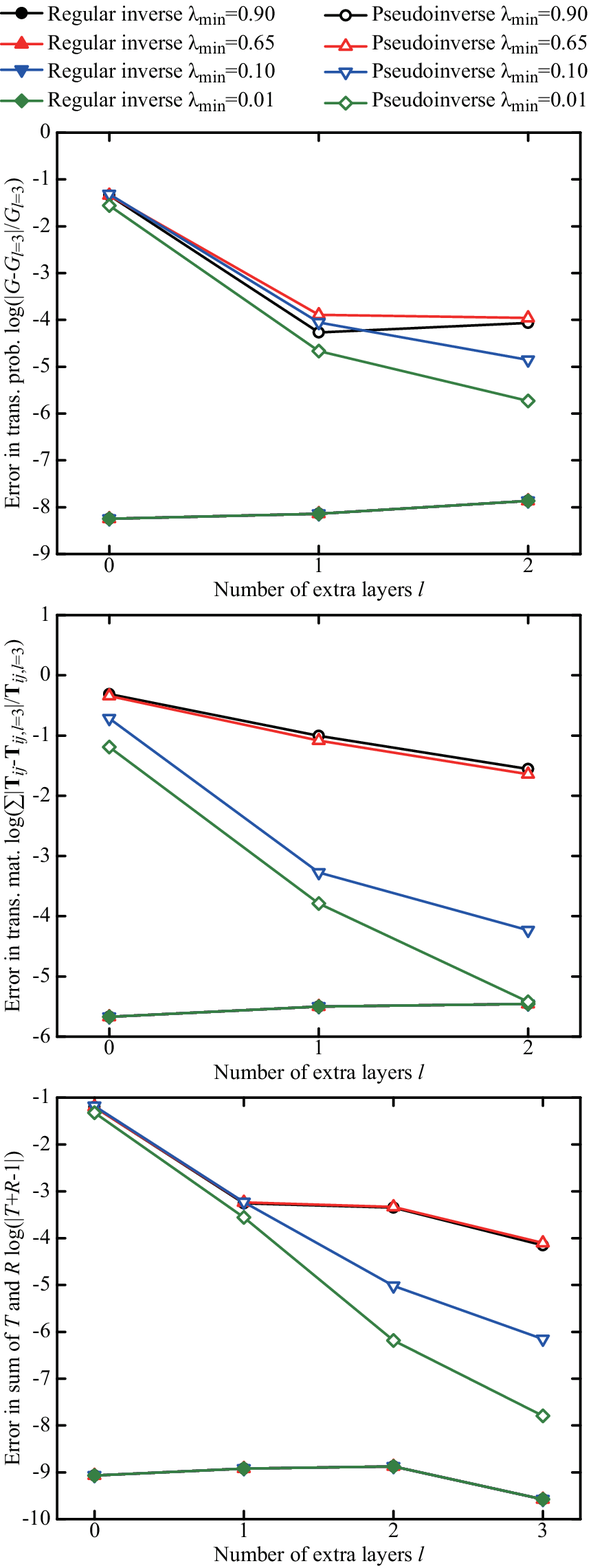}
\caption{(Color online) Difference in transmission probability and matrix and error in the sum of the transmission and reflection probabilities with respect to the number of extra layers $l$. \label{fig:3}}
\end{center}
\end{figure}

\begin{figure}[htb]
\begin{center}
\includegraphics{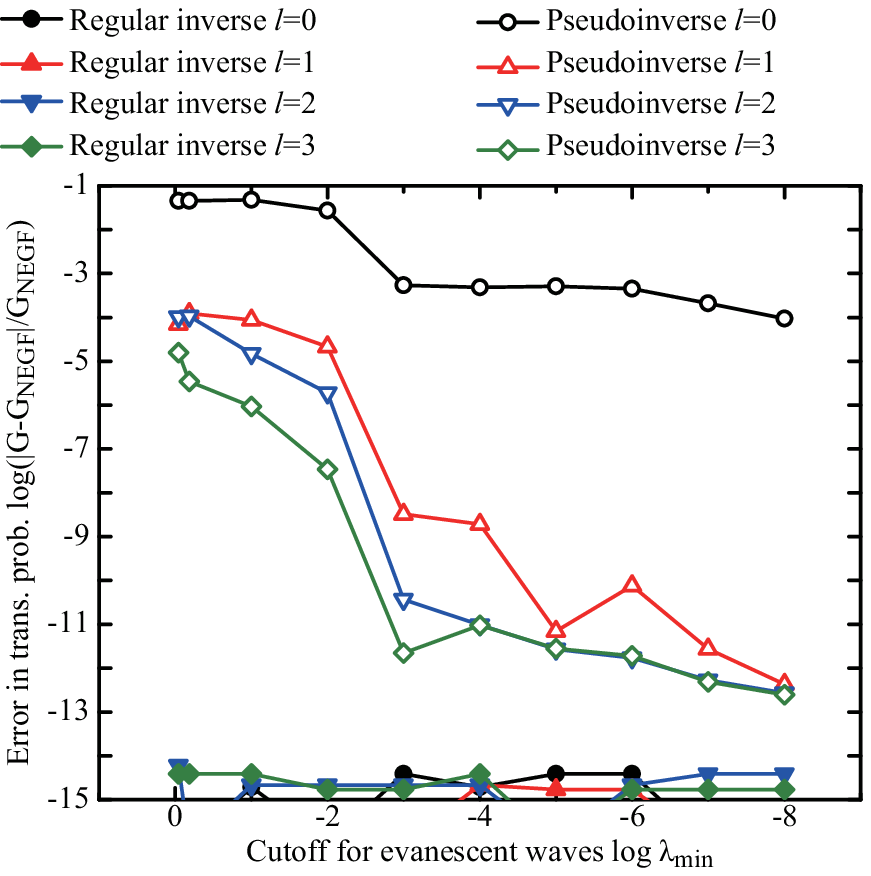}
\caption{(Color online) Difference in the transmission probability from that obtained by the nonequilibrium Green's function method with respect to the cutoff parameter of the evanescent waves $\lambda_{\text{min}}$. \label{fig:4}}
\end{center}
\end{figure}

In Sec.~\ref{sec:Transmission probability}, we described a method for calculating the transmission and reflection probabilities without the pseudoinverses. Here, to demonstrate the accuracy of that method and the degradation due to the pseudoinverses, we examine the variations in the transport properties with respect to the number of extra layers $l$ and cutoff parameter of the evanescent waves $\lambda_{\text{min}}$. The computational model is the same as in Sec.~\ref{sec:Translational invariance of the transmission probability with respect to moving matching plane}. To eliminate the unfavorable effect from the convergence of the ratio matrices on the SWFs, the ratio matrices are converged using the continued-fraction equations Eqs.~(\ref{eqn:06-19}) and (\ref{eqn:06-20}). The behaviors of the transmission probability, transmission matrix, and the sum of the transmission and reflection probabilities with respect to the number of extra layers $l$ are plotted in Fig.~\ref{fig:3}. We can see that the translational invariances of the transmission probability and transmission matrix are well preserved and the sum of the probabilities exactly corresponds with the number of channels in the present method. On the other hand, the transmission probability and transmission matrix probability are clearly affected by the number of extra layers when the pseudoinverses are used. Moreover, the sum of the transmission and reflection probabilities does not correspond to the number of channels, indicating that the pseudoinverses degrade the accuracy. Figure~\ref{fig:4} shows the difference in the transmission probability from that obtained by the nonequilibrium Green's function method with respect to the cutoff parameter $\lambda_{\text{min}}$ of the evanescent waves. We can conclude that the accuracy of the WFM technique on the transmission probability is comparable with that of the nonequilibrium Green's function method even when the rapidly varying evanescent waves ($|\lambda|<\lambda_{\text{min}}$ or $1/\lambda_{\text{min}}<|\lambda|$) are not explicitly computed and the transition region is not extended.

\section{\label{sec:Application}Application}
In this section, we present an application of the ballistic electron transport calculation method discussed in the previous sections to two-dimensional materials composed of graphene sheets. 
Graphene is a well-known two-dimensional (2D) material, and it has been extensively studied for a number of decades.\cite{ MikhailKatsnelson2012,RevModPhys_83_000407,RevModPhys_81_000109,NatureMater_6_000183}
Because of its characteristic electronic band structure, the electron transport properties of graphene sheets, regardless the direction, i.e., in-plane or out-of-plane, continue to attract a great deal of interest. 
Here, we examine in-plane electron transport of graphene sheets with B--N line defects. 

\begin{figure}
\includegraphics{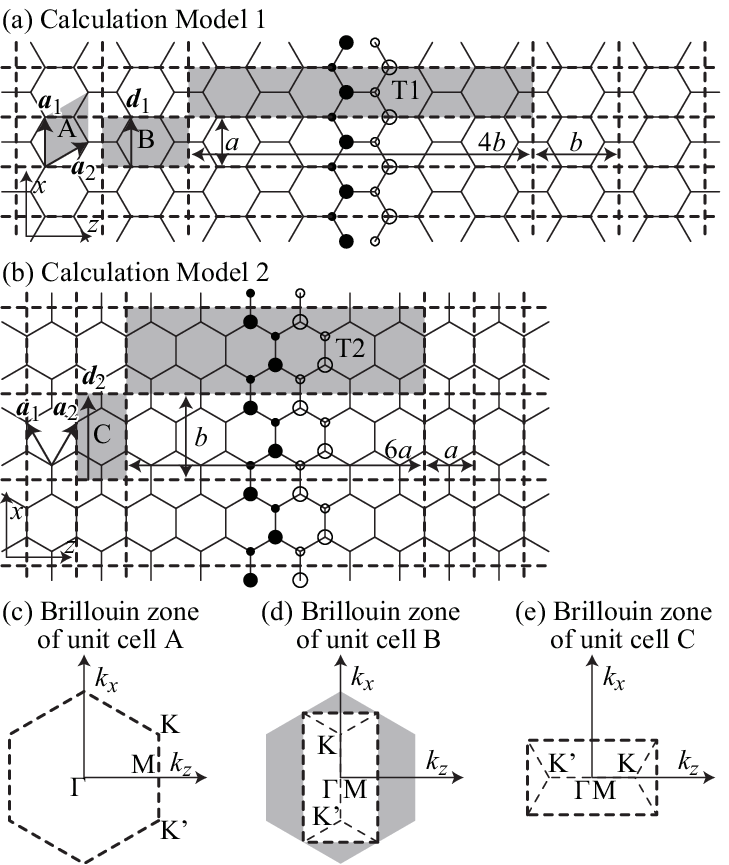}
\caption{\label{fig:FigureA1} Schematic representations of graphene sheets with B--N line defects and Brillouin zones of graphene electrodes. In panels (a) and (b), C atoms are at the corners of the hexagons, and the small solid and open circles denote B atoms and the large ones N atoms. The heteroatoms form a zigzag- or armchair-shaped B--N line defect in each graphene sheet. The dashed lines define the unit cells of the transition regions (T1 and T2) and those of the electrode unit cells (B, and C) with $a=2.46$\AA\ (4.65 $a_{\text{B}}$) and $b=\sqrt{3}a$, while the rhombus labeled A represents the primitive unit cell of a pristine graphene sheet. The vectors $\vec{d}_{1}$ and $\vec{d}_{2}$ denote the translation vectors of the electrode unit cells B and C, respectively. The set of vectors $(\vec{a}_{1},\vec{a}_{2})$ is the primitive translation vectors. Panels (c), (d), and (e) illustrate the first Brillouin zones of the electrode unit cells A, B, and C, respectively. The symbols $\Gamma$, M, K, and K' in panel (c) represent the high-symmetry points, and those in panels (d) and (e) represent the equivalent points after folding the hexagonal Brillouin zone.}
\end{figure}

Figures \ref{fig:FigureA1}(a) and (b) show schematic representations of two different calculation models employed in this study. 
The first model depicted in Fig.~\ref{fig:FigureA1}(a) is referred to as Model 1 hereafter. 
Model 1 has C--C bonds parallel to the $z$ axis and a zigzag-shaped B--N line defect along the $x$ direction. 
On the other hand, the second model, depicted in Fig.~\ref{fig:FigureA1}(b), has C--C bonds perpendicular to the $z$ axis and an armchair-shaped B--N line defect along the $x$ direction. 
We examine two different widths of the line defects for each model, i.e., one is a single-width line defect as indicated by the solid circles in Figs.~\ref{fig:FigureA1}(a) and (b), and the other is a double-width one as indicated by both solid and open circles in Figs.~\ref{fig:FigureA1}(a) and (b). 
The calculation models with the single-width line defect have the suffix ``a'', i.e., Model 1a and Model 2a, and those with the double-width one the suffix ``b'', i.e., Model 1b and Model 2b. 
Although the primitive unit cell of a graphene sheet used for the electrodes is a rhombus containing two carbon atoms [see rhombus A in Fig.~\ref{fig:FigureA1}(a)], we employ a rectangular unit cell containing four carbon atoms as the unit cell of the graphene electrodes [see rectangle B in Fig.~\ref{fig:FigureA1}(a)]. 
The transition region of Model 1 is defined as rectangle T1, and the length in the $z$ direction is four times that of the electrode unit cell B so that the B--N line defect is at the middle of T1, as shown in Fig.~\ref{fig:FigureA1}(a). 
For the other graphene sheet depicted in Fig.~\ref{fig:FigureA1}(b), we define the electrode unit cell as the rectangular four-atom unit cell C and the transition region as rectangle T2. 
The length of T2 in the $z$ direction is six times that of the electrode unit cell C, so that Model 2 has the B--N line defect at the middle of T2, as shown in Fig.~\ref{fig:FigureA1}(b). 

As for the primitive rhombus two-atom unit cell of a graphene sheet, which is indicated by A in Fig.~\ref{fig:FigureA1}(a), it is known that the first Brillouin zone is hexagonal and the high-symmetry K and K' points appear at the corners of the hexagon as represented in Fig.~\ref{fig:FigureA1}(c), where charge carriers are characterized by a linear dispersion in momentum--energy ($\vec{k}$--$E$) space, and the electronic bands form the Dirac cone around the Fermi energy $E_{F}$. 
Here, we use rectangular four-atom unit cells B and C instead of the primitive two-atom unit cells, and thus, the hexagonal Brillouin zone is folded into a rectangular zone, as depicted in Figs.~\ref{fig:FigureA1}(d) and (e). 
As the consequence of the folding, the high-symmetry K and K' points appear inside the rectangular Brillouin zones, as indicated in Figs.~\ref{fig:FigureA1}(d) and (e).

\begin{figure}
\includegraphics{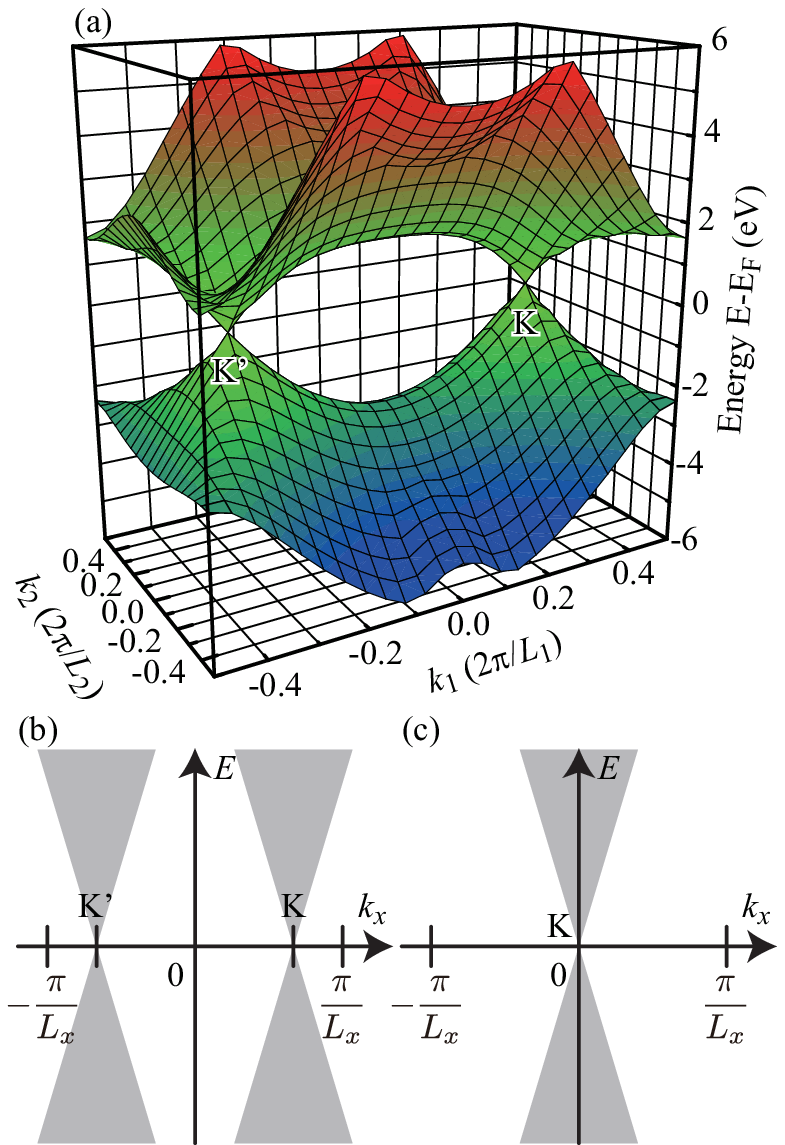}
\caption{\label{fig:FigureA2}(color online) Electronic band structure of a pristine graphene sheet calculated with the rectangular four-atom unit cell, as illustrated in Figs.~\ref{fig:FigureA1}(a) and (b). In panel (a), the axis $k_{1(2)}$ corresponds to the axis $k_{x(y)}$ in Fig.~\ref{fig:FigureA1}(d), and the axis $k_{y(x)}$ in Fig.~\ref{fig:FigureA1}(e). $L_{1}$ and $L_{2}$ denote the lengths of the short and long sides of the rectangular 4-atom unit cell, i.e., $L_{1}=a$ and $L_{2}=b=\sqrt{3}a$, respectively. The Dirac points are labeled K and K'. Panels (b) and (c) depict projections of the band structure on the transversal momentum--energy ($k_{x}$--$E$) plane for Model 1 and 2, respectively. The gray areas are referred to as Dirac triangles in the text.}
\end{figure}

Figure \ref{fig:FigureA2} shows the electronic band structure of a graphene sheet around the Fermi energy $E_{\text{F}}$, as calculated with the rectangular four-atom unit cell. 
The band structure is in good agreement with that calculated by a tight-binding approach with the rectangular four-atom unit cell.\cite{JModPhys_8_000607} 
The Dirac points are clearly visible at the K and K' points in the rectangular Brillouin zone. 
In the energy window between -1 eV and +1 eV, one can see only the electronic states forming the Dirac cone, meaning that the graphene electrodes have conducting electrons with their momentum around the K and K' points within the energy window. 
More specifically, in the case of Model 1, since the Dirac points are at the K and K' points in Fig.~\ref{fig:FigureA1}(d), i.e., $(k_{x},k_{z})=(\pm1/3,0)$,\cite{note1} the electrons conducting through the graphene electrodes and entering T1 have transverse momenta only around $k_{x}=\pm1/3$. 
In the case of Model 2, the Dirac points are at the K and K' points in Fig.~\ref{fig:FigureA1}(e), i.e., $(k_{x},k_{z})=(0,\pm1/3)$, and hence, the conducting electrons of the graphene electrodes have transverse momenta only around $k_{x}=0$. 
Transverse momenta and energies allowed for the electrons conducting through pristine graphene sheets are schematically represented by the gray triangles in Figs.~\ref{fig:FigureA2}(b) and (c), referred to as the Dirac triangles hereafter. 
The Dirac triangles agree well with the classification of graphene transport behavior by Yazyev and Louie. \cite{NatureMater_9_000806} 
According to the classification, Model 1 is class Ib, because the translation vector $\vec{d}_{1}$ [see Fig.~\ref{fig:FigureA1}(a)] has the index $(n,m)=(1,0)$ satisfying $n-m\neq 3q$. On the other hand, Model 2 is class Ia, because the translation vector $\vec{d}_{2}$ [see Fig.~\ref{fig:FigureA1}(b)] has the index $(n,m)=(1,1)$ satisfying $n-m=3q$. 

The electron transport calculations of the graphene sheets with the B--N line defects are carried out using the code \cite{PhysRevE_95_033309,PhysRevB_93_045421,PhysRevE_90_013306,PhysRevB_67_195315,icp} incorporating the aforementioned technique together with the WFM method based on the density functional theory.\cite{RevModPhys_71_001253,PhysRev_136_B864,PhysRev_140_A1133} 
The generalized Bloch wave functions and scattering wave functions are determined in a non-self-consistent manner to a set of given potential and pseudopotential parameters, which are used for constructing the Kohn-Sham matrix $E\mat{S}-\mat{H}$ in Eqs.~\eqref{eqn:01-02} and \eqref{eqn:04-01} for the electrode and transition regions, respectively. 
Electron transmission is determined from the scattering wave functions using Eq.~\eqref{eqn:07-13}. 
The effective potential and pseudopotential parameters are determined in advance under periodic boundary conditions by using the {\sc rspace} electronic structure calculation code,\cite{PhysRevLett_82_005016,PhysRevB_72_085115,PhysRevB_82_205115,icp} which is based on the real-space finite-difference formalism,\cite{PhysRevB_50_011355,PhysRevLett_72_001240} as well as the transport calculation code, in order to treat the physical quantities on the same footing. 
The interaction between the valence electrons and atomic nuclei is treated through the projector-augmented wave pseudopotential method proposed by Bl{\"o}chl,\cite{PhysRevB_50_017953,PhysRevB_59_001758} and the exchange-correlation interaction is calculated using the local density approximation proposed by Vosko, Wilk, and Nasair,\cite{CanJPhys_58_001200} which is within the framework of the density functional theory.\cite{RevModPhys_71_001253,PhysRev_136_B864,PhysRev_140_A1133} 
In electron transport calculations we assume that incident electron waves are injected from the left electrode.

\begin{figure}
\includegraphics{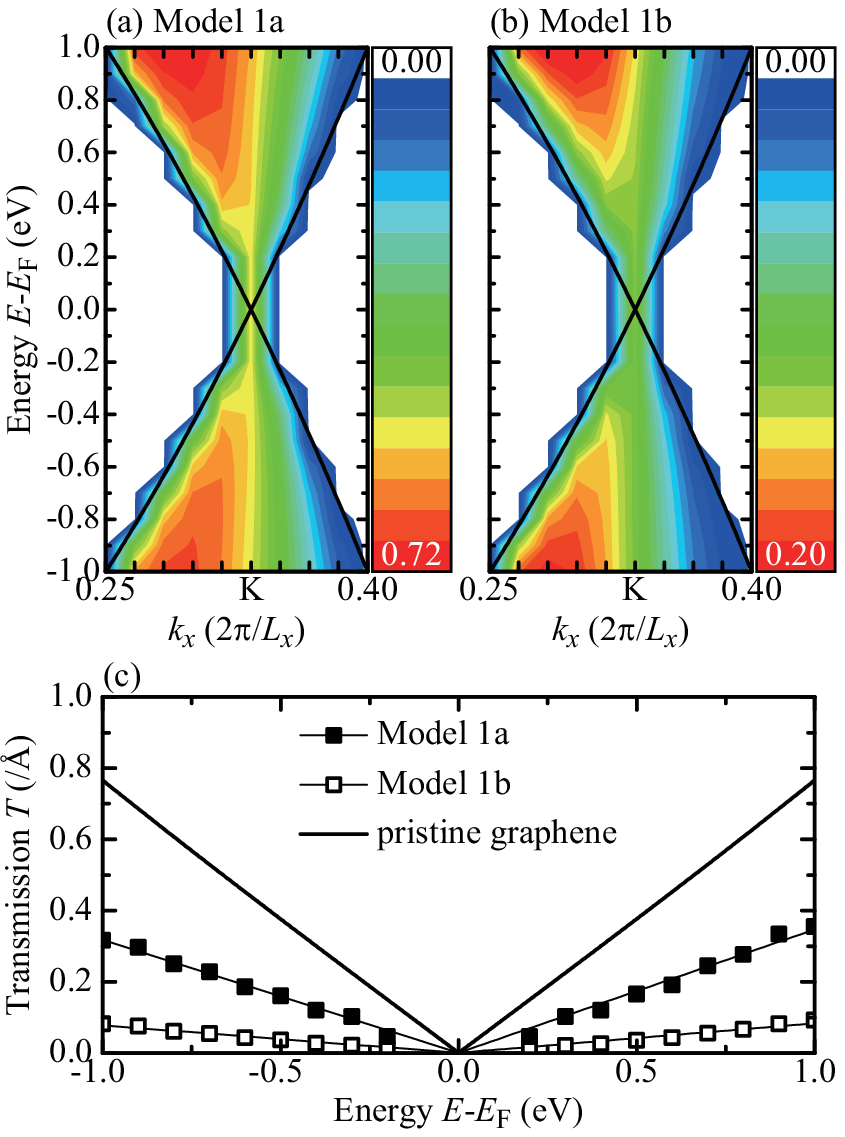}
\caption{\label{fig:FigureA3}(color online) Electron transmissions of graphene sheets with zigzag-shaped B--N line defect [Model 1 in Fig.~\ref{fig:FigureA1}(a)]. In panels (a) and (b), electron transmission contours are plotted as functions of the transverse momentum $k_{x}$ and energy $E$ with respect to the Fermi energy $E_{\text{F}}$ for Model 1a and 1b, respectively. The solid curves describe the Dirac bands of a pristine graphene sheet around the Dirac point K. In panel (c), the transmission spectra per unit length \AA\ for Models 1a and 1b are plotted together with the transmission spectrum of a pristine graphene sheet. The data points indicated by the solid and open squares are evaluated from Eq.~\eqref{eq:EqA2}, and the data points around the Fermi energy $E_{\text{F}}$ are not evaluated because of too few sampling points.}
\end{figure}

Let us move to the discussion on the electron transport properties of the graphene sheets. 
Firstly, we consider electron transport through a pristine graphene sheet without any defects. 
As one can see from Fig.~\ref{fig:FigureA1}(d), the rectangular Brillouin zone of the graphene electrode for Model 1 has Dirac points on the line $k_{z}=0$, and thus, a half of each Dirac cone belong to the longitudinal momentum $k_{z}>0$ and the other half $k_{z}<0$. 
Therefore, only the electrons belonging to the half Dirac cones at one side propagate in the positive $z$ direction, and those belonging to the half Dirac cone at the other side in the negative $z$ direction. 
This means that there is only one incident wave toward the transition region from an electrode for every transverse momentum $k_{x}$ and energy $E$. 
Consequently, when T1 is composed only of a pristine graphene sheet without any heteroatoms, the electron transmission is uniform and quantized to 1 over the Dirac triangles on a $k_{x}$--$E$ plane, as represented in Fig.~\ref{fig:FigureA2}(b). 
Now we discuss the electron transport through the graphene sheets with single and double zigzag-shaped B--N line defects. 
The electron transmission of Models 1a and 1b are shown as functions of the transverse momentum $k_{x}$ and the energy $E$ in Figs.~\ref{fig:FigureA3}(a) and (b), respectively. 
One can easily see that the transmission quantization does not appear when a zigzag-shaped B--N line defect is in T1. 
More specifically, the contour maps show that the electron transmission has a broad and non-quantized peak for the transverse momentum $k_{x}<\text{K}$ and decreases toward the Dirac cone at $K_{x}>\text{K}$. 
This implies that the dependency of the electron transmission on the transverse momentum $k_{x}$ emerges due to the introduction of B--N line defects. 
Comparing the two electron transmission contours, one can see that only the magnitude of the electron transmission changes and the tendencies of the transmission distributions over the $k_{x}$--$E$ plane are almost the same, when the width of the B--N line defect is changed.

Figure \ref{fig:FigureA3}(c) shows the electron transmission per unit length \AA\ for Model 1a, 1b, and a pristine graphene sheet as a function of the energy of incident electrons. 
The electron transmission spectrum $T(E)$ is, in general, evaluated by integrating the total electron transmission function $T_{\text{total}}(k_{x},E)=\sum_{i}T_{i}(k_{x},E)$ over the transverse momentum $k_{x}$, where $ T_{i}(k_{x},E)$ represents the electron transmission of the $i$th transmission channel: 
\begin{align}
T_{\text{BN@Gr}}(E) &= \sum_{i}\int_{-\pi/L_{x}}^{+\pi/L_{x}}\!\!\!\!\!\!T_{i}(k_{x},E)\text{d}k_{x} \\
&\approx \frac{1}{N}\frac{2\pi}{L_{x}}\sum_{i,j}T_{i}(k_{x,j},E). \label{eq:EqA2}
\end{align}
$L_{x}$ is the length of the unit cell in the $x$ direction, and $L_{x}=a$ for Model 1. 
In the practical evaluation, we use Eq.~\eqref{eq:EqA2} in which $j$ indicates a discrete point in the transverse momentum $k_{x}$. 
In the case of a pristine graphene sheet, the total electron transmission is known to be uniformly quantized over the Dirac triangles, and therefore, the electron transmission spectrum is expressed in a simple form using the width of the Dirac triangle in the $k_{x}$ direction, $W_{\text{D}}(E)$, as
\begin{equation}
T_{\text{Gr}}(E)=2W_{\text{D}}(E)\frac{2\pi}{L_{x}}, \label{eq:EqA1}
\end{equation}
where the factor 2 arises from the fact that two Dirac cones centered at K and K' have to be considered. 
One can see from Fig.~\ref{fig:FigureA3}(c) that introducing/widening the B--N line defect drastically decreases the transmission values, though the linearity of the transmission spectra of the pristine graphene sheet is preserved.

\begin{figure}
\includegraphics{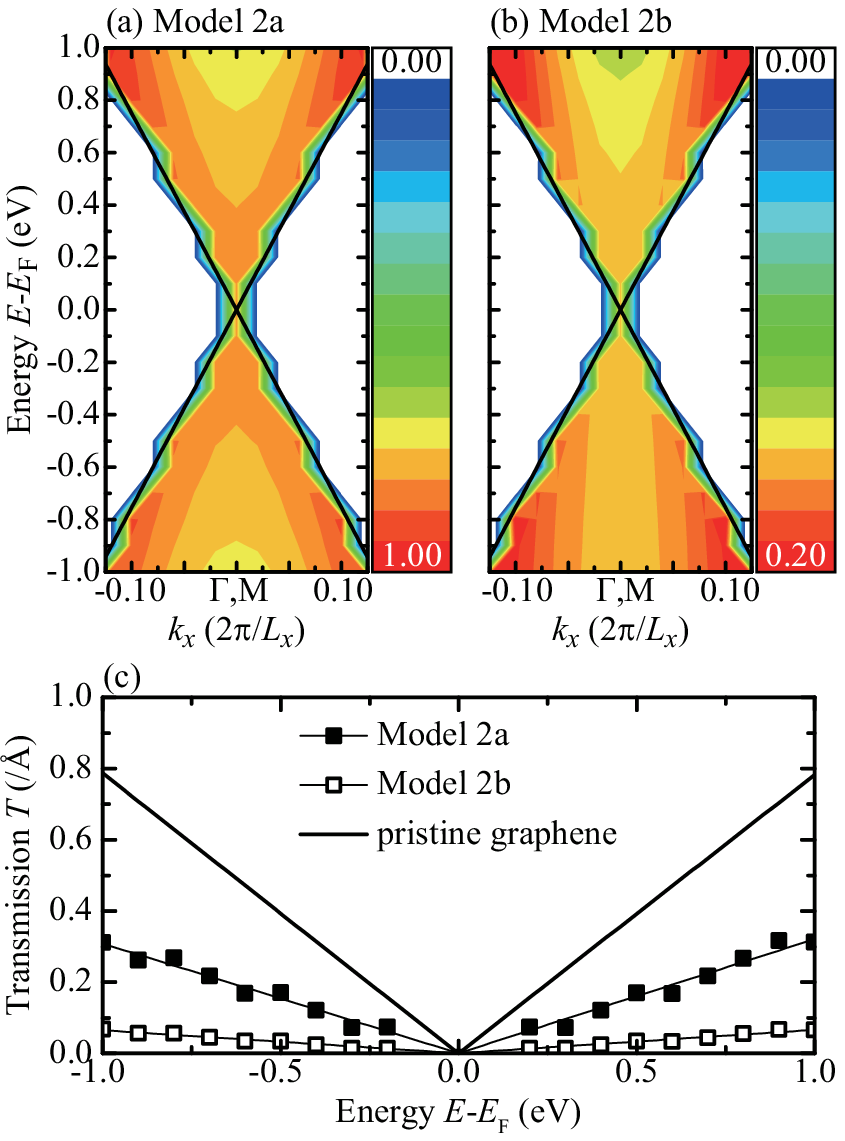}
\caption{\label{fig:FigureA4}(color online) Electron transmissions of the graphene sheets with armchair-shaped B--N line defect [Model 2 in Fig.~\ref{fig:FigureA1}(b)]. In panels (a) and (b), electron transmission contours are plotted as functions of the transverse momentum $k_{x}$ and energy $E$ with respect to the Fermi energy $E_{\text{F}}$ for Models 2a and 2b, respectively. The solid curves describe the Dirac bands of a pristine graphene sheet around the Dirac point K. In panel (c), the transmission spectra per unit length \AA\ for Models 2a and 2b are plotted together with the transmission spectrum of a pristine graphene sheet. The data points indicated by the solid and open squares are evaluated from Eq.~\eqref{eq:EqA2}, and the data points around the Fermi energy $E_{\text{F}}$ are not evaluated because of too few sampling points.}
\end{figure}

In the case of Model 2, the Dirac cones center at $(k_{x},k_{z})=(0,\pm1/3)$ as seen in Fig.~\ref{fig:FigureA1}(e), and overlap to each other on a $k_{x}$--$E$ plane. 
Therefore, there are four electron waves propagating in the $z$ direction for every transverse momentum $k_{x}$ and energy $E$. 
Two of them are heading toward the transition region, and the other two are in the opposite direcion. 
Consequently, the electron transmission of a pristine graphene sheet without any heteroatoms is uniform and quantized to 2 over the Dirac triangles on a $k_{x}$--$E$ plane, as represented in Fig.~\ref{fig:FigureA2}(c). 
The electron transmission of the graphene sheets with the single and double armchair-shaped B--N line defects, i.e., Models 2a and 2b, are shown as functions of the transverse momentum $k_{x}$ and the energy $E$ in Figs.~\ref{fig:FigureA4}(a) and (b), respectively. 
Similar to the case of Model 1, the contour maps exhibit that the electron transmission is drastically reduced by the introduction of the armchair-shaped B--N line defects; the electron transmission values especially around $k_{x}=0$ decrease more than those at around the edges of the Dirac triangles. 
However, the variation in the electron transmission over the Dirac triangles is not as large as in the case of Model 1. 

The electron transmission spectra per unit length \AA\ for Models 2a, 2b, and a pristine graphene sheet are drawn as a function of the energy of incident electrons in Fig.~\ref{fig:FigureA4}(c). 
The electron transmission spectrum for the pristine graphene sheet is evaluated using Eq.~\eqref{eq:EqA1} with $L_{x}=b=\sqrt{3}a$; however, the factor 2 in this case means the two propagating waves entering T2 for every transverse momentum $k_{x}$ and energy $E$. 
The other two transmission spectra for Models 2a and 2b are evaluated from the electron transmission values plotted in Figs.~\ref{fig:FigureA4}(a) and (b) using Eq.~\eqref{eq:EqA2}. 
The electron transmission spectra for Models 2a and 2b are almost linear and there appear to be no transmission peaks and valleys caused by the resonance/antiresonance of incident electron waves with the defect states, as is the case in Models 1a and 1b. 
Comparing the electron transmission spectra for Models 1 and 2, one can suppose that the graphene sheet with the B--N defects has almost isotropic electron transport properties as does the pristine graphene sheet, because the electron transmission spectra for the graphene sheets with the single (double) zigzag- and armchair-shaped B--N defects agree well with each other. 

\begin{figure}
\includegraphics{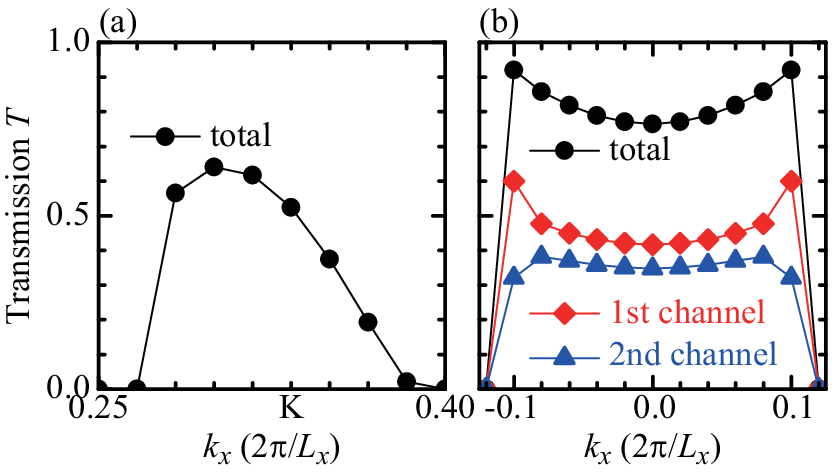}
\caption{\label{fig:FigureA5}(color online) Typical electron transmission profiles of graphene sheets with single zigzag- and armchair-shaped B--N line defects. The electron transmissions are extracted from Fig.~\ref{fig:FigureA3}(a) and \ref{fig:FigureA4}(a) for the energy of -0.8 eV. Panel (b) illustrates the first- and second-channel transmission profiles as well as the total one.}
\end{figure}

Now, let us investigate the difference between Models 1 and 2 as regards the variation in the electron transmission values inside the Dirac triangles, which have been mentioned together with the electron transmission contours in Figs.~\ref{fig:FigureA3} and \ref{fig:FigureA4}. 
Figures~\ref{fig:FigureA5}(a) and (b) illustrate typical electron transmission profiles of Models 1a and 2a extracted from the contour maps in Figs.~\ref{fig:FigureA3}(a) and \ref{fig:FigureA4}(a), respectively. 
The electron transmission of Model 1a rapidly increases to $T=0.64$ as the transverse momentum $k_{x}$ increases from 0.25 to 0.3, while it gradually decreases as the transverse momentum $k_{x}$ changes from 0.3 to 0.4 through the K point. 
On the other hand, the electron transmission of Model 2a rapidly increases up to $T=0.92$ around the transverse momentum $k_{x}=\pm0.1$, and then it gently decreases toward the center $k_{x}=0$. 
This variation in the electron transmission profile for Model 2a can also be seen in the channel-decomposed transmission profiles; in particular, the transmission profile for the second channel preserves the original rectangular-shaped electron transmission profile of the pristine graphene sheet. 
We can say that the electron transmission of Model 1a depends on the transverse momentum $k_{x}$ more significantly than that of Model 2a. 

\begin{figure*}
\includegraphics{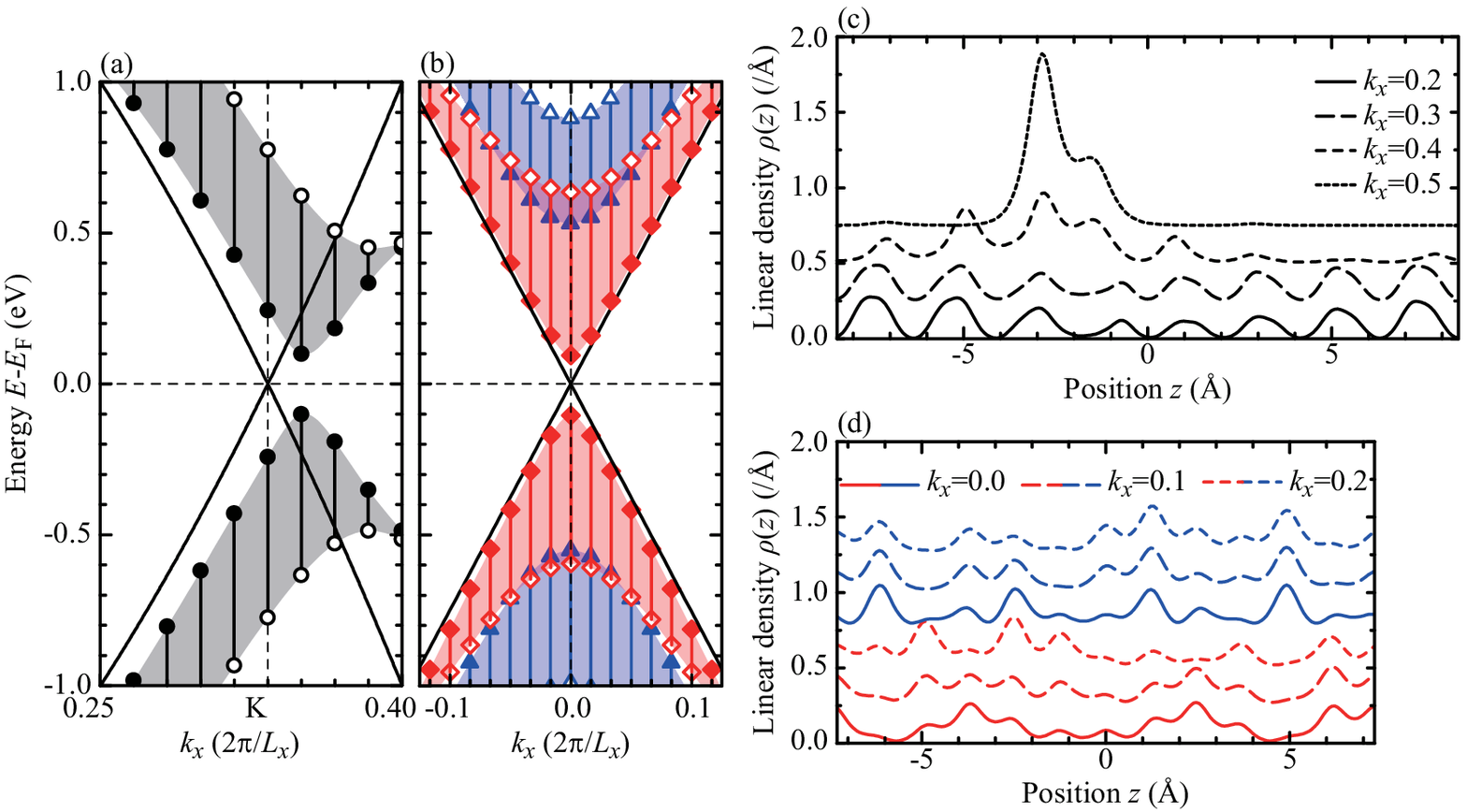}
\caption{\label{fig:FigureA6}(color online) Electronic band structures and spatial distributions of electrons in the bands. The electronic band structures are calculated for the transition regions T1 and T2 in Fig.~\ref{fig:FigureA1} under application of periodic boundary conditions. In panels (a) and (b), the electronic band structures as functions of momentums $k_{x}$ and $k_{z}$ are represented as projections onto the $k_{x}$--$E$ planes for Models 1a and 2a, respectively. The solid and open symbols denote the eigenenergies at $k_{z}=0$ and $k_{z}=0.5$, respectively. The solid curves diagonal to the panels represent the Dirac triangles. Panels (c) and (d) show the spatial distributions of electrons belonging to respective transversal momentum $k_{x}$. In panels (b) and (d), since two electronic bands appear in this energy window, the valence and conduction bands close to the Fermi energy $E_{\text{F}}$ are drawn in red and the others in blue. }
\end{figure*}

Let us discuss the electron-transmission dependence on the transverse momentum $k_{x}$ from the viewpoint of the electronic band structures of the transition regions T1 and T2, which are calculated under periodic boundary conditions. 
Figures~\ref{fig:FigureA6}(a) and (b) depict the electronic band structures projected onto $k_{x}$--$E$ planes for the transition regions T1 and T2 of Models 1a and 2a, respectively. 
From the electronic band structure of Model 1a, one can easily see that the original band structure of the pristine graphene sheet is largely deformed by the introduction of the zigzag-shaped B--N line defect. 
This deformation is attributed to hybridization of the Dirac cone states with defect states, because the spatial distribution of the electronic state changes between localized and delocalized as the transverse momentum $k_{x}$ changes as shown in Fig.~\ref{fig:FigureA6}(c). 
More specifically, the electronic states for the transverse momentums $k_{x}=0.4$ and 0.5 are localized around the defect, while those for $k_{x}=0.2$ and 0.3 are delocalized over the transition region T1. 
Note that the spatial distribution of the $i$th electronic state, $\rho_{i}(z,k_{x})$, is evaluated by
\begin{equation}
\rho_{i}(z,k_{x})=\iiint_{\text{BZ}}|\psi_{i}(x,y,z,k_{x},k_{z})|^2\text{d}k_{z}\text{d}x\text{d}y,
\end{equation}
where $\text{BZ}$ denotes the Brillouin zone and $\psi$ is the Kohn-Sham wave function obtained from the electronic structure calculations under the periodic boundary conditions. 
Consequently, the influence of the localized defect states on the electron transmission depends on the transverse momentum $k_{x}$ in the case of Model 1a. 
On the other hand, in the case of Model 2a, there is no drastic deformation of the band structures caused by the introduction of the armchair-shaped B--N line defect. 
Moreover, in Fig.~\ref{fig:FigureA6}(d), it can be seen that the electronic states are delocalized over the transition region T2 within the transversal-momentum range from $k_{x}=0$ to $k_{x}=0.1$. 
These facts imply that the defect states exist far away from the transverse momentum $k_{x}$ and the energy $E$ presented in Fig.~\ref{fig:FigureA6}(b), and hence, the influence on the electron transmission is less dependent on the transverse momentum $k_{x}$. 
In addition, since no localized defect state exists inside the Dirac triangles shown in Figs.~\ref{fig:FigureA6}(a) and (b), resonant electron transport with a quantized transmission value does not occur in Figs.~\ref{fig:FigureA3}(a) and \ref{fig:FigureA4}(a). 

From the electronic band structures in Figs.~\ref{fig:FigureA6}(a) and (b), one may expect that the electron transmission is forbidden at certain energy ranges, for instance, $E<-0.8$ eV at $k_{x}=\text{K}$ in the case of Model 1a, because of the presence of the band gap. 
This can be explained by imagining an extension of the supercell T1 or T2 in the $z$ direction to separate the interval between the B--N line defects in the neighboring supercells. 
As the supercell length $L_{z}$ increases, the band dispersion of the delocalized states asymptotically approaches that of a pristine graphene sheet, while the localized defect states form dispersionless flat bands in the longitudinal momentum $k_{z}$ direction. 
A consequence of this thought experiment is that electrons with energy $E$ and transverse momentum $k_{x}$ within the Dirac triangles are allowed to enter the graphene sheets with B--N line defects sandwiched between the semi-infinite graphene electrodes and to pass through them with a certain transmission probability, which depends on the influence of the localized defect states, as discussed in the previous paragraph. 

\section{Conclusion}
We reformulated the WFM method for first-principles transport-property calculations so as to exclude the rapidly decreasing evanescent waves because the original WFM methods are formulated to include all the generalized Bloch waves. S\o{}rensen {\it et al.} reported that the transmission probability varies with the increase in the number of extra layers in the transition region and the errors in the transmission and reflection probabilities are not the same. These results indicate that the translational invariance of the transmission probability with respect to moving the matching planes is not preserved and the sum of these probabilities is not equal to the number of channels in the WFM method. We found that the invariance deteriorated as a result of the overlap of the layers between the electrode and transition regions and the pseudoinverses used to exclude the rapidly decreasing evanescent waves, for computing the transmission and reflection coefficients. We proposed a method that eliminates the overlap of the layers between the transition and electrode regions and computes the transmission probability from the SWFs without the pseudoinverses and the Green's function of the transition region. The proposed method nicely recovers the translational invariance of the transmission probability with respect to insertion of the extra layers and gives transmission and reflection probabilities whose sum exactly agrees with the number of channels. We also demonstrated that the accuracy of the transmission probability of the WFM method without computing the rapidly varying evanescent waves or inserting the extra layers in the transition region is comparable with that obtained by the nonequilibrium Green's function method.

We also carried out electron transport calculations on two-dimensional graphene sheets with either zigzag- or armchair-shaped B--N line defects, and discussed about the indirect influence of the defect states on the electron transport properties. 
The transport calculations revealed that the graphene sheet with the zigzag-shaped B--N line defect has an electron transmission dependent on transverse momentum perpendicular to the direction of electron transport. 
On the other hand, the electron transmission of the graphene sheet with the arm-chair-shaped B--N line defect exhibits less dependency on the transverse momentum, though the transmission value is reduced in comparison to that of ideal electron transport without any defects. 
The electron-transmission dependency on transverse momentum can be explained from the electronic band structures of the transition regions calculated under periodic boundary conditions. 
More specifically, the original band structure of a pristine graphene sheet is largely deformed by hybridization with localized states originating from the zigzag-shaped B--N line defect. 
Because the localized states are at the transverse momentum $k_{x}=0.5$, the incident electrons with transverse momenta close to $k_{x}=0.5$ are significantly affected and the transmission value becomes small. 
In the case of the armchair-shaped B--N line defect, the band structure is not significantly deformed, because there are no localized states originating from the B--N line defect at the transverse momentum and energy close to the Dirac points.


\begin{acknowledgments}
S.~T.~and S.~B.~acknowledge the financial support from Deutsche Forschungsgemeinschaft through the Collaborative Research Center SFB 1238 (Project C01). 
T.~O.~acknowledges the financial support from MEXT as a social and scientific priority issue (Creation of new functional devices and high-performance materials to support next-generation industries) to be tackled by using post-K computer and JSPS KAKENHI Grant No.~JP16H03865.
The authors would like to thank Professor Kikuji Hirose of Osaka University and Mr.~Shigeru Iwase of University of Tsukuba for contributing to our discussions.
The numerical calculations were carried out using the supercomputer JURECA at J{\"u}lich Supercomputing Centre, Forschungszentrum J{\"u}lich, the system B of the Institute for Solid State Physics at the University of Tokyo, the COMA of the Center for Computational Sciences at University of Tsukuba, and the K computer provided by the RIKEN Advanced Institute for Computational Science through the HPCI System Research project (Project ID: hp160228).
\end{acknowledgments}

\appendix

\section{Wave function matching for general case}
\label{sec:Wave function matching for general case}
In Sec.~\ref{sec:Wave function matching formula}, we introduced the WFM formula\cite{PhysRevB_67_195315} for the case that the ranks of $\overline{\mat{H}}_{L}(\overline{\mat{H}}_{R})$ and $\overline{\mat{H}}_{L,L}(\overline{\mat{H}}_{R,R})$ are identical ($m_L = m_{LL}$ and $m_R = m_{RR}$). In many cases, e.g., real-space grid methods, the Hamiltonian does not satisfy such a condition. In this subsection, the WFM formula for the case that $m_L > m_{LL}$ and $m_R > m_{RR}$ will be derived. Although we assume $m_L > 2m_{LL}$, this condition can be easily satisfied by increasing the number of unit cells in $\overline{\mat{H}}_{L}$, as introduced in Sec.~\ref{sec:Generalized Bloch waves in electrodes}. The splitting using $\xi_1$, $\xi_2$, and $\xi_3$ in Sec.~\ref{sec:Generalized Bloch waves in electrodes} allows the SWFs $\vec{\psi}_1^{ref,(l)}$ to be written as
\begin{eqnarray}
\label{eqn:06-01}
\vec{\psi}_1^{ref,(l)}=
\left(
\begin{array}{c}
\vec{\psi}_1^{ref,(l)}(\xi_1) \\
\vec{\psi}_1^{ref,(l)}(\xi_2) \\
\vec{\psi}_1^{ref,(l)}(\xi_3)
\end{array}
\right).
\end{eqnarray}
Using the same separation for $\vec{\psi}_0$, $\vec{\psi}_n^{(l)}$, and $\vec{\psi}_{n+1}$, Eq.~(\ref{eqn:04-01}) is expressed as
\begin{equation}
\label{eqn:06-02}
E\widehat{\mat{S}}^{(l)}-\widehat{\mat{H}}^{(l)}
\left(
\begin{array}{c}
\vec{\psi}_1^{ref,(l)}(\xi_1) \\
\vec{\psi}_1^{ref,(l)}(\xi_2) \\
\vec{\psi}_1^{ref,(l)}(\xi_3) \\
\vdots \\
\vec{\psi}_n^{(l)}(\xi_1) \\
\vec{\psi}_n^{(l)}(\xi_2) \\
\vec{\psi}_n^{(l)}(\xi_3)
\end{array}
\right)
=
\left(
\begin{array}{c}
-\overline{\mat{h}}^{L \dag}\vec{\psi}_{0}(\xi_3) \\
0 \\
\vdots \\
0 \\
-\overline{\mat{h}}^R\vec{\psi}_{n+1}(\xi_1)
\end{array}
\right).
\end{equation}
Multiplying $\widehat{\mat{G}}^{(l)}[=(E\widehat{\mat{S}}^{(l)}-\widehat{\mat{H}}^{(l)})^{-1}]$ from the left hand side leads
\begin{equation}
\label{eqn:06-03}
\left(
\begin{array}{c}
\vec{\psi}_1^{ref,(l)}(\xi_1) \\
\vec{\psi}_1^{ref,(l)}(\xi_2) \\
\vec{\psi}_1^{ref,(l)}(\xi_3) \\
\vdots \\
\vec{\psi}_n^{(l)}(\xi_1) \\
\vec{\psi}_n^{(l)}(\xi_2) \\
\vec{\psi}_n^{(l)}(\xi_3)
\end{array}
\right)
=
\widehat{\mat{G}}^{(l)}\left(
\begin{array}{c}
-\overline{\mat{h}}^{L \dag}\vec{\psi}_{0}(\xi_3) \\
0 \\
\vdots \\
0 \\
-\overline{\mat{h}}^R\vec{\psi}_{n+1}(\xi_1)
\end{array}
\right).
\end{equation}

Supposing that $( i, j)$ block-matrix element of Green's function $\mat{G}^{(l)}_{i,j}$ in Eq.~(\ref{eqn:04-03}) is given by
\begin{eqnarray}
\label{eqn:06-04}
\mat{G}^{(l)}_{i,j}=
\left(
\begin{array}{ccc}
\mat{G}^{(l)}_{i,j}(\xi_1,\xi_1) & \mat{G}^{(l)}_{i,j}(\xi_1,\xi_2) & \mat{G}^{(l)}_{i,j}(\xi_1,\xi_3) \\
\mat{G}^{(l)}_{i,j}(\xi_2,\xi_1) & \mat{G}^{(l)}_{i,j}(\xi_2,\xi_2) & \mat{G}^{(l)}_{i,j}(\xi_2,\xi_3) \\
\mat{G}^{(l)}_{i,j}(\xi_3,\xi_1) & \mat{G}^{(l)}_{i,j}(\xi_3,\xi_2) & \mat{G}^{(l)}_{i,j}(\xi_3,\xi_3)
\end{array}
\right),
\end{eqnarray}
we have the WFM formula,
\begin{eqnarray}
\label{eqn:06-05}
\left(
\begin{array}{cc}
- \mat{G}^{(l)}_{1-l,1-l}(\xi_1,\xi_1)\overline{\mat{h}}^{L \dag} - \vec{\psi}_{1}^{ref,(l)}(\xi_1)\left( \vec{\psi}_{0}^{ref}(\xi_3) \right)^{-1} & - \mat{G}^{(l)}_{1-l,n+l}(\xi_1,\xi_3)\overline{\mat{h}}^{R} \\
- \mat{G}^{(l)}_{n+l,1-l}(\xi_3,\xi_1)\overline{\mat{h}}^{L \dag} & - \mat{G}^{(l)}_{n+l,n+l}(\xi_3,\xi_3)\overline{\mat{h}}^{R} - \vec{\psi}_{n}^{(l)}(\xi_3)\left( \vec{\psi}_{n+1}(\xi_1) \right)^{-1}
\end{array}
\right) \times \nonumber \\
\left(
\begin{array}{c}
\vec{\psi}_{0}^{ref}(\xi_3) \\
\vec{\psi}_{n+1}(\xi_1)
\end{array}
\right)
=
\left(
\begin{array}{c}
\mat{G}^{(l)}_{1-l,1-l}(\xi_1,\xi_1)\overline{\mat{h}}^{L \dag} \vec{\psi}_{0}^{in}(\xi_3) + \vec{\psi}_{1}^{in,(l)}(\xi_1) \\
\mat{G}^{(l)}_{n+l,1-l}(\xi_3,\xi_1)\overline{\mat{h}}^{L \dag} \vec{\psi}_{0}^{in}(\xi_3)
\end{array}
\right). \nonumber \\
\end{eqnarray}
Applying the same separation using $\xi_1$, $\xi_2$, and $\xi_3$ to $\tilde{\mat{\Phi}}_{R}^{+}$ and $\tilde{\mat{\Phi}}_{L}^{-}$, Eqs.~(\ref{eqn:04-09}) and (\ref{eqn:04-10}) take the forms of 
\begin{eqnarray}
\label{eqn:06-06}
\vec{\psi}_{n+1}(\xi_3)\left( \vec{\psi}_{n+2}(\xi_1) \right)^{-1} &=&
\tilde{\mat{\Phi}}_{R}^{+}(\xi_3)
\left( \tilde{\mat{\Lambda}}_{R}^{+}\tilde{\mat{\Phi}}_{R}^{+}(\xi_1) \right)^{-1}, \\
\label{eqn:06-07}
\vec{\psi}_{0}^{ref}(\xi_1)\left( \vec{\psi}_{-1}^{ref}(\xi_3) \right)^{-1} &=& \tilde{\mat{\Phi}}_{L}^{-}(\xi_1)
\left( (\tilde{\mat{\Lambda}}_{L}^{-})^{-1}\tilde{\mat{\Phi}}_{L}^{-}(\xi_3) \right)^{-1},
\end{eqnarray}
respectively. Since the rapidly decreasing evanescent waves are not included in $\tilde{\mat{\Lambda}}_{R}^{+}\tilde{\mat{\Phi}}_{R}^{+}(\xi_1)$ and $(\tilde{\mat{\Lambda}}_{L}^{-})^{-1}\tilde{\mat{\Phi}}_{L}^{-}(\xi_3)$, $\left( \tilde{\mat{\Lambda}}_{R}^{+}\tilde{\mat{\Phi}}_{R}^{+}(\xi_1) \right)^{-1}$ and $\left( (\tilde{\mat{\Lambda}}_{L}^{-})^{-1}\tilde{\mat{\Phi}}_{L}^{-}(\xi_3) \right)^{-1}$ are computed by the pseudoinverses.
The ratio of the SWFs in Eq.~(\ref{eqn:06-05}) is an $m_{LL} \times m_{LL}$ ($m_{RR} \times m_{RR}$) matrix, while that in Eq.~(\ref{eqn:04-07}) is an $m_L \times m_L$ ($m_R \times m_R$) matrix. It is not possible to solve Eqs.~(\ref{eqn:04-12}) and (\ref{eqn:04-13}) using $\vec{\psi}_{0}^{ref}(\xi_1)(\vec{\psi}_{-1}^{ref}(\xi_3))^{-1}$ and $\vec{\psi}_{n+1}(\xi_3)(\vec{\psi}_{n+2}(\xi_1))^{-1}$ strictly because $m_L > m_{LL}$ and $m_R > m_{RR}$. We derive other continued-fraction equations for the ratio of SWFs. $\vec{\psi}_{1}^{ref,(l)}$ obeys the equation,
\begin{eqnarray}
\label{eqn:06-09}
\left(
\begin{array}{c}
\vec{\psi}_{0}^{ref}(\xi_1) \\
\vec{\psi}_{0}^{ref}(\xi_2) \\
\vec{\psi}_{0}^{ref}(\xi_3) \\
\end{array}
\right)=\left(
\begin{array}{ccc}
\bm{\thetaup}_{L}(\xi_1,\xi_1) & \bm{\thetaup}_{L}(\xi_1,\xi_2) & \bm{\thetaup}_{L}(\xi_1,\xi_3) \\
\bm{\thetaup}_{L}(\xi_2,\xi_1) & \bm{\thetaup}_{L}(\xi_2,\xi_2) & \bm{\thetaup}_{L}(\xi_2,\xi_3) \\
\bm{\thetaup}_{L}(\xi_3,\xi_1) & \bm{\thetaup}_{L}(\xi_3,\xi_2) & \bm{\thetaup}_{L}(\xi_3,\xi_3)
\end{array}
\right)
\left(
\begin{array}{c}
-\overline{\mat{h}}^{L \dag}\vec{\psi}_{-1}^{ref}(\xi_3) \\
0 \\
-\overline{\mat{h}}^{L}\vec{\psi}_{1}^{ref,(l)}(\xi_1) \\
\end{array}
\right),
\end{eqnarray}
where $\bm{\thetaup}_{L}(\xi_i,\xi_j)$ is the $( i, j)$th block-matrix element of $(\overline{\mat{H}}_L)^{-1}$ defined in Eq.~(\ref{eqn:01-04}). From the first and third block rows of Eq.~(\ref{eqn:06-09}), we see that
\begin{eqnarray}
\label{eqn:06-10}
\vec{\psi}_{0}^{ref}(\xi_1) &=& - \bm{\thetaup}_{L}(\xi_1,\xi_1)\overline{\mat{h}}^{L \dag}\vec{\psi}_{-1}^{ref}(\xi_3) - \bm{\thetaup}_{L}(\xi_1,\xi_3)\overline{\mat{h}}^{L}\vec{\psi}_{1}^{ref,(l)}(\xi_1), \\
\label{eqn:06-11}
\vec{\psi}_{0}^{ref}(\xi_3) &=& - \bm{\thetaup}_{L}(\xi_3,\xi_1)\overline{\mat{h}}^{L \dag}\vec{\psi}_{-1}^{ref}(\xi_3) - \bm{\thetaup}_{L}(\xi_3,\xi_3)\overline{\mat{h}}^{L}\vec{\psi}_{1}^{ref,(l)}(\xi_1).
\end{eqnarray}
Multiplying $(\vec{\psi}_{-1}^{ref}(\xi_3))^{-1}$ from the right side of Eq.~(\ref{eqn:06-10}) leads to
\begin{equation}
\label{eqn:06-12}
\vec{\psi}_{0}^{ref}(\xi_1)\left( \vec{\psi}_{-1}^{ref}(\xi_3) \right)^{-1} = - \bm{\thetaup}_{L}(\xi_1,\xi_1)\overline{\mat{h}}^{L \dag} - \bm{\thetaup}_{L}(\xi_1,\xi_3)\overline{\mat{h}}^{L}\vec{\psi}_{1}^{ref,(l)}(\xi_1)\left( \vec{\psi}_{-1}^{ref}(\xi_3) \right)^{-1}.
\end{equation}
Multiplying $(\vec{\psi}_{1}^{ref,(l)}(\xi_1))^{-1}$ from the right side of Eq.~(\ref{eqn:06-11}), we have
\begin{equation}
\label{eqn:06-13}
\vec{\psi}_{0}^{ref}(\xi_3)\left( \vec{\psi}_{1}^{ref,(l)}(\xi_1) \right)^{-1} = - \bm{\thetaup}_{L}(\xi_3,\xi_1)\overline{\mat{h}}^{L \dag}\vec{\psi}_{-1}^{ref}(\xi_3)\left( \vec{\psi}_{1}^{ref,(l)}(\xi_1) \right)^{-1} - \bm{\thetaup}_{L}(\xi_3,\xi_3)\overline{\mat{h}}^{L}.
\end{equation}
Substituting $\vec{\psi}_{1}^{ref,(l)}(\xi_1)(\vec{\psi}_{-1}^{ref}(\xi_3))^{-1}$  in Eq.~(\ref{eqn:06-12}) into $\vec{\psi}_{-1}^{ref}(\xi_3)(\vec{\psi}_{1}^{ref,(l)}(\xi_1))^{-1}$  in Eq.~(\ref{eqn:06-13}), we obtain the continued-fraction equation for the left electrode side,
\begin{eqnarray}
\label{eqn:06-14}
\vec{\psi}_{0}^{ref}(\xi_3)\left( \vec{\psi}_{1}^{ref,(l)}(\xi_1) \right)^{-1} &=& - \bm{\thetaup}_{L}(\xi_3,\xi_3)\overline{\mat{h}}^{L} + \bm{\thetaup}_{L}(\xi_3,\xi_1)\overline{\mat{h}}^{L \dag} \times\nonumber \\
&& \left( \vec{\psi}_{0}^{ref}(\xi_1) \left( \vec{\psi}_{-1}^{ref}(\xi_3) \right)^{-1} + \bm{\thetaup}_{L}(\xi_1,\xi_1)\overline{\mat{h}}^{L \dag} \right)^{-1}\bm{\thetaup}_{L}(\xi_1,\xi_3)\overline{\mat{h}}^{L}.
\end{eqnarray}
Analogous to Eq.~(\ref{eqn:06-14}), the continued-fraction equation for the right electrode side is given by
\begin{eqnarray}
\label{eqn:06-15}
\vec{\psi}_{n+1}(\xi_1)\left( \vec{\psi}_{n}^{(l)}(\xi_3) \right)^{-1} &=& - \bm{\thetaup}_{R}(\xi_1,\xi_1)\overline{\mat{h}}^{R \dag} + \bm{\thetaup}_{R}(\xi_1,\xi_3)\overline{\mat{h}}^{R} \times\nonumber \\
&& \left( \vec{\psi}_{n+1}(\xi_3)\left( \vec{\psi}_{n+2}(\xi_1) \right)^{-1} + \bm{\thetaup}_{R}(\xi_3,\xi_3)\overline{\mat{h}}^{R} \right)^{-1}\bm{\thetaup}_{R}(\xi_3,\xi_1)\overline{\mat{h}}^{R \dag}.
\end{eqnarray}

Shifting the matching planes by one extra layer inside of the transition region, we obtain the following WFM formula in which the left (right) matching plane is set between $\vec{\psi}_{1}^{(l)}(\xi_3)$ and $\vec{\psi}_{1}^{(l-1)}(\xi_1)$ ($\vec{\psi}_{n}^{(l)}(\xi_1)$ and $\vec{\psi}_{n}^{(l-1)}(\xi_3)$).
\begin{eqnarray}
\label{eqn:06-08}
\left(
\begin{array}{cc}
- \mat{G}^{(l-1)}_{2-l,2-l}(\xi_1,\xi_1)\overline{\mat{h}}^{L \dag} - \vec{\psi}_{1}^{ref,(l-1)}(\xi_1)\left( \vec{\psi}_{1}^{ref,(l)}(\xi_3) \right)^{-1} & - \mat{G}^{(l-1)}_{2-l,n+l-1}(\xi_1,\xi_3)\overline{\mat{h}}^{R} \\
- \mat{G}^{(l-1)}_{n+l-1,2-l}(\xi_3,\xi_1)\overline{\mat{h}}^{L \dag} & - \mat{G}^{(l-1)}_{n+l-1,n+l-1}(\xi_3,\xi_3)\overline{\mat{h}}^{R} - \vec{\psi}_{n}^{(l-1)}(\xi_3)\left( \vec{\psi}_{n}^{(l)}(\xi_1) \right)^{-1}
\end{array}
\right) \times \nonumber \\
\left(
\begin{array}{c}
\vec{\psi}_{1}^{ref,(l)}(\xi_3) \\
\vec{\psi}_{n}^{(l)}(\xi_1)
\end{array}
\right)
=
\left(
\begin{array}{c}
\mat{G}^{(l-1)}_{2-l,2-l}(\xi_1,\xi_1)\overline{\mat{h}}^{L \dag} \vec{\psi}_{1}^{in,(l)}(\xi_3) + \vec{\psi}_{1}^{in,(l-1)}(\xi_1) \\
\mat{G}^{(l-1)}_{n+l-1,2-l}(\xi_3,\xi_1)\overline{\mat{h}}^{L \dag} \vec{\psi}_{1}^{in,(l)}(\xi_3)
\end{array}
\right). \nonumber \\
\end{eqnarray}
By repeatedly solving Eq.~(\ref{eqn:06-15}) $l$ times, the WFM formula is rewritten as
\begin{eqnarray}
\label{eqn:06-16}
\left(
\begin{array}{cc}
- \mat{G}^{(0)}_{1,1}(\xi_1,\xi_1)\overline{\mat{h}}^{L \dag} - \vec{\psi}_{1}^{ref,(0)}(\xi_1)\left( \vec{\psi}_{1}^{ref,(1)}(\xi_3) \right)^{-1} & - \mat{G}^{(0)}_{1,n}(\xi_1,\xi_3)\overline{\mat{h}}^{R} \\
- \mat{G}^{(0)}_{n,1}(\xi_3,\xi_1)\overline{\mat{h}}^{L \dag} & - \mat{G}^{(0)}_{n,n}(\xi_3,\xi_3)\overline{\mat{h}}^{R} - \vec{\psi}_{n}^{(0)}(\xi_3)\left( \vec{\psi}_{n}^{(1)}(\xi_1)\right)^{-1}
\end{array}
\right) \times \nonumber \\
\left(
\begin{array}{c}
\vec{\psi}_{1}^{ref,(1)}(\xi_3) \\
\vec{\psi}_{n}^{(1)}(\xi_1)
\end{array}
\right)
=
\left(
\begin{array}{c}
\mat{G}^{(0)}_{1,1}(\xi_1,\xi_1)\overline{\mat{h}}^{L \dag} \vec{\psi}_{1}^{in,(1)}(\xi_3) + \vec{\psi}_{1}^{in,(0)}(\xi_1) \\
\mat{G}^{(0)}_{n,1}(\xi_3,\xi_1)\overline{\mat{h}}^{L \dag} \vec{\psi}_{1}^{in,(1)}(\xi_3)
\end{array}
\right).
\end{eqnarray}
Defining the $m_{LL} \times m_{LL}$ and $m_{RR} \times m_{RR}$ ratio matrices for the left and right electrode sides, 
\begin{eqnarray}
\label{eqn:06-17}
\mat{R}^{ref,(l-1)} &=& \vec{\psi}_{1}^{ref,(l)}(\xi_3)\left( \vec{\psi}_{1}^{ref,(l-1)}(\xi_1) \right)^{-1}, \\
\label{eqn:06-18}
\mat{R}^{tra,(l)} &=& \vec{\psi}_{n}^{(l)}(\xi_1)\left( \vec{\psi}_{n}^{(l-1)}(\xi_3)\right)^{-1},
\end{eqnarray}
respectively, Eqs.~(\ref{eqn:06-14}), (\ref{eqn:06-15}), and (\ref{eqn:06-16}) are expressed as
\begin{eqnarray}
\label{eqn:06-19}
\mat{R}^{ref,(l-1)} = - \bm{\thetaup}_{L}(\xi_3,\xi_3)\overline{\mat{h}}^{L} + \bm{\thetaup}_{L}(\xi_3,\xi_1)\overline{\mat{h}}^{L \dag} \left( \left( \mat{R}^{ref,(l)} \right)^{-1} + \bm{\thetaup}_{L}(\xi_1,\xi_1)\overline{\mat{h}}^{L \dag} \right)^{-1}\bm{\thetaup}_{L}(\xi_1,\xi_3)\overline{\mat{h}}^{L}, \\
\label{eqn:06-20}
\mat{R}^{tra,(l)} = - \bm{\thetaup}_{R}(\xi_1,\xi_1)\overline{\mat{h}}^{R \dag} + \bm{\thetaup}_{R}(\xi_1,\xi_3)\overline{\mat{h}}^{R} \left( \left( \mat{R}^{tra,(l+1)} \right)^{-1} + \bm{\thetaup}_{R}(\xi_3,\xi_3)\overline{\mat{h}}^{R} \right)^{-1}\bm{\thetaup}_{R}(\xi_3,\xi_1)\overline{\mat{h}}^{R \dag},
\end{eqnarray}
\begin{eqnarray}
\label{eqn:06-21}
\left(
\begin{array}{cc}
- \mat{G}^{(0)}_{1,1}(\xi_1,\xi_1)\overline{\mat{h}}^{L \dag} - \left( \mat{R}^{ref,(0)} \right)^{-1} & - \mat{G}^{(0)}_{1,n}(\xi_1,\xi_3)\overline{\mat{h}}^{R} \\
- \mat{G}^{(0)}_{n,1}(\xi_3,\xi_1)\overline{\mat{h}}^{L \dag} & - \mat{G}^{(0)}_{n,n}(\xi_3,\xi_3)\overline{\mat{h}}^{R} - \left( \mat{R}^{tra,(1)} \right)^{-1}
\end{array}
\right)
\left(
\begin{array}{c}
\vec{\psi}_{1}^{ref,(1)}(\xi_3) \\
\vec{\psi}_{n}^{(1)}(\xi_1)
\end{array}
\right) \nonumber \\
=
\left(
\begin{array}{c}
\mat{G}^{(0)}_{1,1}(\xi_1,\xi_1)\overline{\mat{h}}^{L \dag} \vec{\psi}_{1}^{in,(1)}(\xi_3) + \vec{\psi}_{1}^{in,(0)}(\xi_1) \\
\mat{G}^{(0)}_{n,1}(\xi_3,\xi_1)\overline{\mat{h}}^{L \dag} \vec{\psi}_{1}^{in,(1)}(\xi_3)
\end{array}
\right).
\end{eqnarray}
From the discussion in Sec.~\ref{sec:Moving matching plane of wave function matching formula}, the ratio matrices are uniquely determined if $l$ is sufficiently large. Inserting $\vec{\psi}_{1}^{ref,(1)}(\xi_3)$ and $\vec{\psi}_{n}^{(1)}(\xi_1)$ into Eq.~(\ref{eqn:06-03}), the SWFs in the transition region are computed.

The self-energy terms are defined by
\begin{eqnarray}
\label{eqn:06-22}
\mat{\Sigma}_L^{(l)} &=& -\overline{\mat{h}}^{L \dag} \mat{R}^{ref,(l)}, \\
\label{eqn:06-23}
\mat{\Sigma}_R^{(l)} &=& -\overline{\mat{h}}^{R} \mat{R}^{tra,(l+1)}.
\end{eqnarray}
Then, the similar equation with Eq.~(\ref{eqn:05-10}) becomes
\begin{eqnarray}
\label{eqn:06-25}
E\widehat{\mat{S}}^{(0)}-\widehat{\mat{H}}^{(0)}-\widetilde{\mat{H}}^{(0)}
\left(
\begin{array}{c}
\vec{\psi}_{1}^{(0)}(\xi_1) \\
\vdots \\
\vdots \\
\vdots \\
\vec{\psi}_{n}^{(0)}(\xi_3)
\end{array}
\right)
=
\left(
\begin{array}{c}
-\overline{\mat{h}}^{L \dag}\vec{\psi}_{1}^{in,(1)}(\xi_3) - \mat{\Sigma}_L^{(0)} \vec{\psi}_{1}^{in,(0)}(\xi_1) \\
0 \\
\vdots \\
0 \\
0
\end{array}
\right),
\end{eqnarray}
where
\begin{eqnarray}
\label{eqn:06-26}
\widetilde{\mat{H}}^{(0)}= \left(
\begin{array}{ccccc}
\mat{\Sigma}_L^{(0)} & 0 & \cdots & \cdots & 0 \\
0 & 0 & & & \vdots \\
\vdots & & \ddots & & \vdots \\
\vdots & & & 0 & 0 \\
0 & \cdots & \cdots & 0 & \mat{\Sigma}_R^{(0)} \\
\end{array}
\right).
\end{eqnarray}
Here $\mat{\Sigma}_L^{(0)}$ and $\mat{\Sigma}_R^{(0)}$ are $m_{LL} \times m_{LL}$ and $m_{RR} \times m_{RR}$ block matrices. The relations of Eq.~(\ref{eqn:04-26}) and (\ref{eqn:04-27}) give the SWFs in the electrode side of the matching planes.

For the computation of the transmission probability, the matrices containing the group velocities of the incident and transmitted waves, and the vector of the transmission coefficients are given by
\begin{eqnarray}
\label{eqn:07-14}
\mat{v}^{in} &=& ia\mat{\Phi}_L^{+ \dag}(\xi_1) \mat{\Gamma}_L^{(0)} \mat{\Phi}_L^+(\xi_1), \\
\label{eqn:07-15}
\check{\mat{v}}^{tra} &=& ia(\check{\mat{\Phi}}_{R}^+)^\dag(\xi_1) \mat{\Gamma}_{R}^{(0)} \check{\mat{\Phi}}_{R}^+(\xi_1), \\
\label{eqn:07-16}
\check{\vec{t}}_k &=& \left( \check{\mat{\Phi}}_{R}^{+}(\xi_1) \right)^{-1}\vec{\psi}_{n,k}^{(1)}(\xi_1),
\end{eqnarray}
respectively. The transmission matrix is obtained by
\begin{equation}
\label{eqn:07-17}
T_{k,k}=ia\left( \mat{v}_{k,k}^{in} \right)^{-1} \vec{\psi}_{n,k}^{(1) \dag}(\xi_1) \mat{\Gamma}_{R}^{(0)} \vec{\psi}_{n,k}^{(1)}(\xi_1)
\end{equation}
with the dimension of $\mat{\Gamma}_{R}^{(0)}$ being $m_{RR}$.

\section{Invariance of the transmission matrix with respect to the transmitted waves} 
\label{sec:Invariance of the transmission matrix with respect to the transmission waves}
We prove Eq.~(\ref{eqn:07-12}) indicating that the transmission matrix is invariant as long as a regular matrix is taken for $\check{\mat{\Phi}}_{R}^{+}$. Let the number of columns of $\widetilde{\mat{\Phi}}_{R}^{+}$ ($\mathring{\mat{\Phi}}_{R}^{+}$) is $\widetilde{m}_R$ ($\mathring{m}_R$). Defining a $m_R \times m_R$ regular matrix
\begin{eqnarray}
\label{eqn:08-01}
\mat{P}=
\left(
\begin{array}{cc}
\mat{I} & \mat{p}_1 \\
\mat{0} & \mat{p}_2 \\
\end{array}
\right)
\end{eqnarray}
with $\mat{p}_1$ and $\mat{p}_2$ being $\widetilde{m}_R \times \mathring{m}_R$ and $\mathring{m}_R \times \mathring{m}_R$ matrices, respectively, $\mat{\Phi}_{R}^{+}$ and $\check{\mat{\Phi}}_{R}^{+}$ are related as
\begin{equation}
\label{eqn:08-02}
\mat{\Phi}_{R}^{+} = \check{\mat{\Phi}}_{R}^{+} \mat{P}.
\end{equation}
From Eq.~(\ref{eqn:08-01}), one sees that $\check{\mat{\Phi}}_{R}^{+}$ can include the components of $\widetilde{\mat{\Phi}}_{R}^{+}$. Equation~(\ref{eqn:07-10}) is rewritten as
\begin{eqnarray}
\label{eqn:08-03}
\check{\vec{t}}_k &=& \left( \check{\mat{\Phi}}_{R}^{+} \right)^{-1}\vec{\psi}_{n,k}^{(1)} \nonumber \\
&=& \mat{P} \left( \mat{\Phi}_{R}^{+} \right)^{-1}\vec{\psi}_{n,k}^{(1)} \nonumber \\
&=& \mat{P} \vec{t}_k.
\end{eqnarray}
Then, we have
\begin{equation}
\label{eqn:08-04}
\check{\mat{t}}=\mat{P} \mat{t}.
\end{equation}
The group velocity of the transmitted wave is expressed as
\begin{eqnarray}
\label{eqn:08-05}
\check{\mat{v}}^{tra} &=& ia \left( \check{\mat{\Phi}}_{R}^{+} \right)^\dag \mat{\Gamma}_R^{(0)} \check{\mat{\Phi}}_{R}^{+} \nonumber \\
&=& ia \left( \mat{\Phi}_{R}^{+} \mat{P}^{-1} \right)^\dag \mat{\Gamma}_R^{(0)} \mat{\Phi}_{R}^{+} \mat{P}^{-1} \nonumber \\
&=& \left( \mat{P}^{-1} \right)^\dag \mat{v}^{tra} \mat{P}^{-1}.
\end{eqnarray}
Substituting Eqs.~(\ref{eqn:08-02}) and (\ref{eqn:08-03}) into Eq.~(\ref{eqn:07-01}) leads to
\begin{eqnarray}
\label{eqn:08-06}
\mat{T}&=&\left( \mat{v}^{in} \right)^{-1} \check{\mat{t}}^\dag \check{\mat{v}}^{tra} \check{\mat{t}} \nonumber \\
&=&\left( \mat{v}^{in} \right)^{-1} \left( \mat{P} \mat{t} \right)^\dag \left( \mat{P}^{-1} \right)^\dag \mat{v}^{tra} \mat{P}^{-1} \mat{P} \mat{t} \nonumber \\
&=&\left( \mat{v}^{in} \right)^{-1}\mat{t}^\dag \mat{v}^{tra} \mat{t}.
\end{eqnarray}

\section{Deterioration of the translational invariance of the transmission probability due to the pseudoinverse}
\label{sec:Deterioration of the translational invariance of the transmission probability due to the pseudoinverse}
Although we proved that the transmission matrix is invariant with respect to the elements of $\check{\mat{\Phi}}_{R}^{+}$ in Appendix~\ref{sec:Invariance of the transmission matrix with respect to the transmission waves}, the transmission coefficient should not change except the trivial Bloch factors as long as the same matrices for the group velocity are used to retain the translational invariance of the transmission probability. The transmission coefficients at the first and second extra layers from the boundary between the right electrode and transition regions, $\widetilde{\vec{t}}^{(l)}_{n,k}$ and $\widetilde{\vec{t}}^{(l-1)}_{n,k}$, should satisfy the following relation,
\begin{equation}
\label{eqn:b-01}
\widetilde{\vec{t}}^{(l)}_{n,k}=\widetilde{\mat{\Lambda}}_R^+ \widetilde{\vec{t}}^{(l-1)}_{n,k}.
\end{equation}
According to Eqs.~(\ref{eqn:03-04}) and (\ref{eqn:03-06}), the SWFs at the first and second extra layers are described as
\begin{eqnarray}
\label{eqn:b-02}
\vec{\psi}_{n,k}^{(l)}&=&\left(
\begin{array}{c}
\widetilde{\mat{\Phi}}_{R}^{+}, \mathring{\mat{\Phi}}_{R}^{+}
\end{array}
\right)\left(
\begin{array}{c}
(\widetilde{\mat{\Lambda}}_{R}^{+})^{-1}\vec{\tilde{a}}_{n+1,k}^{+} \\
(\mathring{\mat{\Lambda}}_{R}^{+})^{-1}\vec{\mathring{a}}_{n+1,k}^{+}
\end{array}
\right), \\
\label{eqn:b-03}
\vec{\psi}_{n,k}^{(l-1)}&=&\left(
\begin{array}{c}
\widetilde{\mat{\Phi}}_{R}^{+}, \mathring{\mat{\Phi}}_{R}^{+}
\end{array}
\right)\left(
\begin{array}{c}
(\widetilde{\mat{\Lambda}}_{R}^{+})^{-2}\vec{\tilde{a}}_{n+1,k}^{+} \\
(\mathring{\mat{\Lambda}}_{R}^{+})^{-2}\vec{\mathring{a}}_{n+1,k}^{+}
\end{array}
\right),
\end{eqnarray}
respectively. Note that in the transition region, the evanescent waves excluded by the cutoff parameter $\lambda_{\text{min}}$ also contribute to the SWFs. Multiplying the pseudoinverse $(\widetilde{\mat{\Phi}}_{R}^{+})^{-1}$ to Eq.~(\ref{eqn:b-03}) from left hand side leads to
\begin{eqnarray}
\label{eqn:b-04}
\widetilde{\vec{t}}^{(l)}_{n,k}&=& \left( \widetilde{\mat{\Phi}}_{R}^{+} \right)^{-1}\left(
\begin{array}{c}
\widetilde{\mat{\Phi}}_{R}^{+}, \mathring{\mat{\Phi}}_{R}^{+}
\end{array}
\right)\left(
\begin{array}{c}
(\widetilde{\mat{\Lambda}}_{R}^{+})^{-1}\vec{\tilde{a}}_{n+1,k}^{+} \\
(\mathring{\mat{\Lambda}}_{R}^{+})^{-1}\vec{\mathring{a}}_{n+1,k}^{+}
\end{array}
\right) \nonumber \\
&=& (\widetilde{\mat{\Lambda}}_{R}^{+})^{-1}\vec{\tilde{a}}_{n+1,k}^{+} + \left( \widetilde{\mat{\Phi}}_{R}^{+} \right)^{-1}\mathring{\mat{\Phi}}_{R}^{+}(\mathring{\mat{\Lambda}}_{R}^{+})^{-1}\vec{\mathring{a}}_{n+1,k}^{+}, \\
\widetilde{\vec{t}}^{(l-1)}_{n,k}&=& \left( \widetilde{\mat{\Phi}}_{R}^{+} \right)^{-1}\left(
\label{eqn:b-05}
\begin{array}{c}
\widetilde{\mat{\Phi}}_{R}^{+}, \mathring{\mat{\Phi}}_{R}^{+}
\end{array}
\right)\left(
\begin{array}{c}
(\widetilde{\mat{\Lambda}}_{R}^{+})^{-2}\vec{\tilde{a}}_{n+1,k}^{+} \\
(\mathring{\mat{\Lambda}}_{R}^{+})^{-2}\vec{\mathring{a}}_{n+1,k}^{+}
\end{array}
\right) \nonumber \\
&=& (\widetilde{\mat{\Lambda}}_{R}^{+})^{-2}\vec{\tilde{a}}_{n+1,k}^{+} + \left( \widetilde{\mat{\Phi}}_{R}^{+} \right)^{-1}\mathring{\mat{\Phi}}_{R}^{+}(\mathring{\mat{\Lambda}}_{R}^{+})^{-2}\vec{\mathring{a}}_{n+1,k}^{+}.
\end{eqnarray}
Since the generalized Bloch waves are nonorthogonal, $( \widetilde{\mat{\Phi}}_{R}^{+})^{-1}\mathring{\mat{\Phi}}_{R}^{+} \neq \mat{0}$, resulting in $\widetilde{\vec{t}}^{(l)}_{n,k} \neq \widetilde{\mat{\Lambda}}_R^+ \widetilde{\vec{t}}^{(l-1)}_{n,k}$. When a regular matrix is used for $\check{\mat{\Phi}}_{R}^{+}$, the second terms of Eqs.~(\ref{eqn:b-04}) and (\ref{eqn:b-05}) vanish and the translational invariance is preserved.

\section{Convergence behavior of the ratio matrix}
\label{sec:Convergence behavior of the ratio matrix}
\begin{figure}[htb]
\begin{center}
\includegraphics{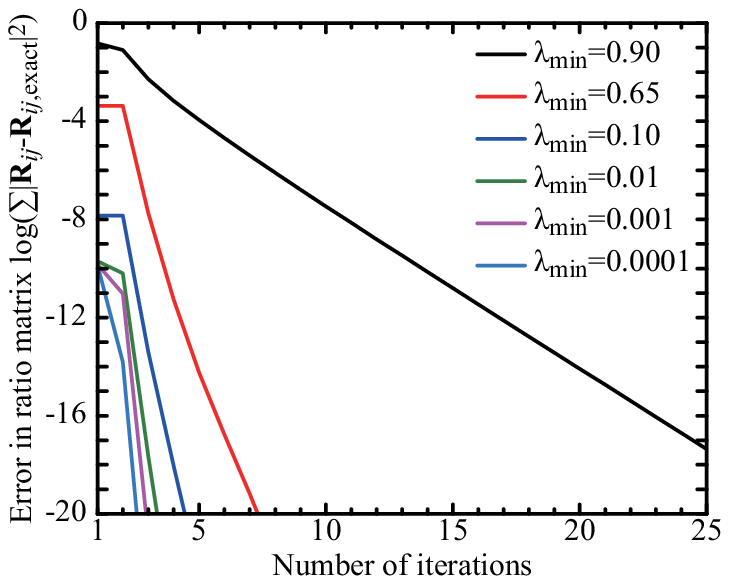}
\caption{(Color online) Convergence behavior of the ratio matrix $\mat{R}^{ref}$. \label{fig:5}}
\end{center}
\end{figure}

As introduced in Sec.~\ref{sec:Wave function matching formula}, the ratio matrices $\mat{R}^{ref,(0)}$ and $\mat{R}^{tra,(1)}$ are obtained by solving the continued-fraction equations Eqs.~(\ref{eqn:04-20}) and (\ref{eqn:04-21}). The initial matrix of $(\tilde{\mat{\Lambda}}_{L}^{-})^{-1}\tilde{\mat{\Phi}}_{L}^{-}( \tilde{\mat{\Phi}}_{L}^{-} )^{-1}$ ($(\tilde{\mat{\Lambda}}_{R}^{+})^{-1}\tilde{\mat{\Phi}}_{R}^{+}( \tilde{\mat{\Phi}}_{R}^{+} )^{-1}$) is constructed at $\vec{\psi}_{-1}$ and $\vec{\psi}_{0}$ ($\vec{\psi}_{n+1}$ and $\vec{\psi}_{n+2}$), in which the rapidly decreasing evanescent waves vanish. It is of importance that the ratio matrix $\mat{R}^{ref,(0)}$ ($\mat{R}^{tra,(1)}$) is uniquely determined after the continued-fraction equation Eq.~(\ref{eqn:04-20}) (Eq.~(\ref{eqn:04-21})) is solved until $\mat{R}^{ref,(0)}$ ($\mat{R}^{tra,(1)}$) on the left- and right-hand sides become identical. Figure~\ref{fig:5} shows the convergence behavior of the ratio matrix $\mat{R}^{ref}$ for the (9,0) CNT electrode used in Sec.~\ref{sec:Translational invariance of the transmission probability with respect to moving matching plane} as the electrode region. $\mat{R}^{ref}_{exact}$ are computed by solving Eq.~(\ref{eqn:06-19}) 10 times using the initial value of $(\tilde{\mat{\Lambda}}_{L}^{-})^{-1}\tilde{\mat{\Phi}}_{L}^{-}( \tilde{\mat{\Phi}}_{L}^{-} )^{-1}$ with $\lambda_{\text{min}}=10^{-10}$ as a reference and the difference from $\mat{R}^{ref}_{exact}$ are plotted in Fig.~\ref{fig:5}. We can see that the number of iterations for solving the continued-fraction equations is small when $\lambda_{\text{min}}$ is set to be small and the ratio matrices can be uniquely determined by the continued-fraction equations.

\begin{thebibliography}{99}
\bibitem{note1}
Throughout this section, the wave number $k_{x(z)}$ is normalized by $2\pi/L_{x(z)}$, where $L_{x(z)}$ denotes the length of the transition region in the $x(z)$ direction.
\end{thebibliography}
\end{document}